\documentclass[a4paper, 11pt]{article}
\pdfoutput=1
\usepackage{jcappub}
\usepackage{afterpage}
\usepackage{multirow}
\usepackage{rotating}
\usepackage{epstopdf}
\usepackage{amsmath}
\usepackage{amsthm}
\usepackage{mathrsfs}
\usepackage{subfigure}

\title{Primordial black holes induced stochastic axion-photon oscillations in primordial magnetic field}

\author{Hai-Jun Li}
\affiliation{Center for Advanced Quantum Studies, Department of Physics, Beijing Normal University, Beijing 100875, China}
\emailAdd{lihaijun@bnu.edu.cn}

\abstract{
Primordial black holes (PBHs) can be produced in the very early Universe due to the large density fluctuations.
The cosmic background of axion-like particles (ALPs) could be non-thermally generated by PBHs.
In this paper, we investigate the ALPs emitted by ultra-light PBHs with the mass range $10 \, {\rm g} \lesssim M_{\rm PBH} \lesssim 10^9 \, \rm g$, in which PBHs would have completely evaporated before the start of Big Bang Nucleosynthesis (BBN) and can therefore not be directly constrained.
In this case, the minimal scenario that ALPs could interact only with photons is supposed.
We study the stochastic oscillations between the ALPs and photons in the cosmic magnetic field in detail.
The primordial magnetic field (PMF) can be modelled as the stochastic background field model with the completely non-homogeneous component of the cosmic plasma.
Using the latest stringent limits on PMF, we show the numerical results of ALP-photon oscillation probability distributions with the homogeneous and stochastic magnetic field scenarios.
The PBH-induced stochastic ALP-photon oscillations in the PMF may have the effects on some further phenomena, such as the cosmic microwave background (CMB), the cosmic X-ray background (CXB), and the extragalactic gamma-ray background (EGB).
}

\arxivnumber{2208.04605}

\begin{document}
\maketitle

\section{Introduction}

Primordial black holes (PBHs) have recently gathered significant interest as the candidates of dark matter (DM) \cite{Bird:2016dcv}.
They are a kind of black holes that may be produced due to the large density fluctuations ($\rm i.e.$, the radiation pressure could not resist the gravitational collapse in the overdense spherically symmetric regions surrounded by the compensating void) in the very early Universe \cite{Hawking:1971ei, Carr:1974nx, Carr:1975qj}, see $\rm e.g.$ refs.~\cite{Carr:2020gox, Carr:2020xqk, Carr:2021bzv} for recent reviews.  

PBHs could exist over a wide range of mass scales from the Planck mass to $\sim \mathcal{O}(10^{10})$ solar
masses \cite{Carr:2016drx}. 
They emit the energetic particles through Hawking evaporation.
PBHs larger than $M_{\rm PBH} \sim \mathcal{O}(10^{15})\, \rm g$ may have not entirely evaporated by now, which could account for a significant fraction of the cold DM \cite{Chapline:1975ojl}.
There are several constraints on the abundance of non-evaporated PBHs, such as the evaporation constraints \cite{Carr:2009jm, Strong:2004ry}, the lensing constraints \cite{Barnacka:2012bm, Niikura:2019kqi}, the dynamical constraints \cite{Carr:1997cn, Koushiappas:2017chw}, the cosmic structure constraints \cite{Carr:2018rid, Afshordi:2003zb}, and the gravitational wave constraints \cite{Raidal:2017mfl, Hooper:2020evu}, etc.
On the contrary, those PBHs with the mass range less than $M_{\rm PBH} \sim \mathcal{O}(10^{15})\, \rm g$ may have completely evaporated at present, while it could also lead the observable signals on the cosmological consequences.
There are also various constraints on the evaporating PBHs, such as the Big Bang Nucleosynthesis (BBN) constraints \cite{Kohri:1999ex, Carr:2009jm}, the cosmic microwave background (CMB) constraints \cite{Ricotti:2007au, Ali-Haimoud:2016mbv, Cang:2020aoo}, the PBH explosion constraints \cite{Maki:1995pa, Linton:2006yu}, and the extragalactic (Galactic) gamma-ray background (EGB) and cosmic-ray constraints \cite{Page:1976wx, Carr:2016hva, MacGibbon:1991vc, Bambi:2008kx, Lunardini:2019zob, Cang:2021owu}, etc.

Additionally, the cosmic axion(-like) background (C$a$B) could be non-thermally generated by PBHs \cite{Dror:2021nyr}.
One of the leading hypotheses is that ultralight spin-0 bosons such as axion or axion-like particle (ALP) are potential DM candidates if they are non-thermally produced in the early Universe by misalignment mechanism \cite{Preskill:1982cy, Sikivie:2009fv, Marsh:2015xka}.
In refs.~\cite{Bernal:2021yyb, Bernal:2021bbv, Choi:2022btl}, they recently investigated the phenomenological consequences of axion and ALP as the DM triggered by PBHs.
Axion is originally proposed with the spontaneous breaking of the global $U(1)$ PQ symmetry to solve the strong CP problem in quantum chromodynamics (QCD) \cite{Peccei:1977ur, Peccei:1977hh, Weinberg:1977ma, Wilczek:1977pj}, which is also defined as the QCD axion.
While ALP is predicted by the extensions of Standard Model (SM) in the high energy, such as the string theory \cite{Svrcek:2006yi, Arvanitaki:2009fg}, without having to solve the strong CP problem.
See $\rm e.g.$ refs.~\cite{DiLuzio:2020wdo, Choi:2020rgn, Galanti:2022ijh} for recent reviews of QCD axion and ALP.  
ALP is the ultralight pseudo-Nambu-Goldstone boson (pNGB) $a$ with the two-photon vertex
\begin{eqnarray}
\mathscr{L}_{\rm }\supset-\frac{1}{4}g_{a\gamma}aF_{\mu\nu}\tilde{F}^{\mu\nu}\, ,
\end{eqnarray}
where $g_{a\gamma}$ is the coupling constant of ALP-photon. 
The ALP-photon oscillation in the astrophysical magnetic fields would lead to several detectable effects, such as the TeV transparency of the Universe \cite{Hooper:2007bq, Li:2020pcn, Li:2021zms, Li:2022jgi}, the anomalous stellar cooling problem \cite{Giannotti:2015kwo, Giannotti:2017hny}, the galaxy cluster center soft X-ray excess \cite{Conlon:2013txa, Angus:2013sua}, and the XENON1T low energy electronic recoil event excess \cite{XENON:2020rca, DiLuzio:2020jjp}, etc.\footnote{The continuously updated ALP-photon limits can be found in \cite{ciaran_o_hare_2020_3932430}.}

Magnetic field is ubiquitous in the Universe, which would appear wherever plasmas and currents can be found.
The primordial magnetic field (PMF) might have been generated with the high temperature environment of the Big Bang \cite{Subramanian:2015lua}, which also has the effects on various cosmological phenomena, such as the inflation \cite{Turner:1987bw, Ratra:1991bn}, the cosmological phase transitions \cite{Vachaspati:1991nm, Grasso:1997nx, Brandenburg:2017neh}, the BBN \cite{Grasso:1994ph, Kahniashvili:2010wm}, the CMB temperature and polarization
anisotropies \cite{Jedamzik:1999bm, Kunze:2013uja}, the large scale structure (LSS) formation \cite{Kim:1994zh, Battaner:1996jk}, and the cosmic gravity wave background \cite{Caprini:2001nb, Caprini:2006jb, RoperPol:2019wvy}, see $\rm e.g.$ refs.~\cite{Yamazaki:2012pg, Durrer:2013pga,Subramanian:2015lua} for recent reviews.
The most widely used PMF model is the stochastic background field model with the non-helical and helical components.
The latest upper limits on PMF are derived by the Planck collaboration using the Planck 2015 data \cite{Planck:2015zrl}, with the magnetic field strength $B \sim \mathcal{O}(1)\, \rm nG$ for the scale $\lambda \sim \mathcal{O}(1)\, \rm Mpc$.
We also note that the very recent stringent limits on PMF with $B \sim 8.9-47\, \rm pG$ \cite{Jedamzik:2018itu} derived from the CMB radiation.
 
In this work, we are more concerned about the phenomenon of ALP-photon coupling induced by PBHs in the PMF.
We note that the impacts of ALP-photon oscillations triggered by PBHs on cosmic X-ray background (CXB) and X-ray excess in galaxy clusters are recently investigated in ref.~\cite{Schiavone:2021imu}.
Similar to refs.~\cite{Evoli:2016zhj, Schiavone:2021imu}, we focus our attention on the ALPs emitted by PBHs with the mass range \cite{Papanikolaou:2020qtd}
\begin{eqnarray}
\mathcal{O}(10) \, {\rm g} \lesssim M_{\rm PBH} \lesssim \mathcal{O}(10^9)  \, \rm g \, ,
\end{eqnarray}
in which PBHs would have completely evaporated before the start of BBN.
In this case, the minimal scenario that ALPs could interact only with photons is supposed.
Differently from \cite{Schiavone:2021imu}, we just study the stochastic oscillations between ALPs and photons induced by PBHs in the cosmic magnetic field, but not intend to investigate its phenomenological consequences in this context.
The PMF model and ALP-photon oscillation in the magnetic field are discussed in detail.
We consider the PMF model with the stochastic background field, which can be modelled as the completely non-homogeneous component of the cosmic plasma.
We show the ALP-photon oscillation probability in different cases for comparisons.
In addition to the CXB, the ALP-photon oscillations induced by PBHs may also have the effects on CMB and EGB, etc.

The rest of this paper is structured as follows.
In section~\ref{sec_PBH}, we introduce the PBH formation and evaporation, including the Schwarzschild black holes and Kerr black holes, and a brief description of ALPs emitted by PBHs.
In section~\ref{sec_PMF}, we describe the PMF model and the latest limits on PMF.
Section~\ref{sec_oscillations} is the ALP-photon oscillations in the PMF, including the general ALP-photon oscillation in the magnetic field and the numerical results.
Our numerical results of the ALP-photon oscillation probability in the homogeneous and stochastic magnetic fields are also presented in this section.
Finally, we comment on our results and conclude in section~\ref{sec_conclusions}.

\section{Primordial black hole formation and evaporation}
\label{sec_PBH}

PBHs could be generated due to large density perturbations in the very early Universe \cite{Hawking:1971ei}.
When these density fluctuations reenter the horizon, they can collapse to form a black hole if they exceed a threshold \cite{Carr:1974nx, Carr:1975qj}.
In this paper, we consider the formation of PBHs in the radiation dominated epoch.
The PBH initial mass at the formation time $t_{\rm in}$ can be described by \cite{Carr:2009jm}
\begin{eqnarray}
M_{\rm PBH} = \frac{4\pi}{3}\frac{\gamma \rho_R(t_{\rm in})}{H^3(t_{\rm in})}\, ,
\end{eqnarray}
with the gravitational collapse factor $\gamma \sim \mathcal{O}(1)$ in the radiation dominated epoch, the SM radiation energy density $\rho_R$, and the Hubble expansion rate $H$.
Using the Friedmann equation
\begin{eqnarray}
H=\sqrt{\frac{\rho_R}{3M_{\rm Pl}^2}} \, ,  
\end{eqnarray}
we can derive 
\begin{eqnarray}
M_{\rm PBH} = \frac{4\pi M_{\rm Pl}^2}{H(t_{\rm in})}\, ,
\end{eqnarray}
where $M_{\rm Pl}$ ($\equiv 1/\sqrt{8\pi G}$) is the reduced Planck mass.

After formation, PBH evaporates by releasing all possible degrees of freedom in nature, which is also called Hawking evaporation.
The Hawking evaporation time of PBH can be described by \cite{Hawking:1974rv}
\begin{eqnarray}
t_{\rm ev}=\frac{160}{\pi g_*}\frac{M_{\rm PBH}^3}{M_{\rm Pl}^4}\, ,
\end{eqnarray}
where $g_*\sim \mathcal{O}(100)$ is the number of relativistic degrees of freedom contributing to the SM energy density.
Considering that PBHs have completely evaporated before the start of BBN, we have
\begin{eqnarray}
H(t_{\rm ev}) > H(t_{\rm BBN})\, .
\end{eqnarray}
With $H(t_{\rm ev}) = 1/(2t_{\rm ev})$, $H(t_{\rm BBN})=\sqrt{\rho_{\rm BBN}/(3M_{\rm Pl}^2)}$, and $\rho_{\rm BBN}^{1/4} \sim 1\, \rm MeV$, we can derive $M_{\rm PBH} \lesssim 10^9 \, \rm g \,$.
On the other hand, the PBHs should be formed after the end of inflation, then  we have
\begin{eqnarray}
H(t_{\rm in}) < H(t_{\rm inf})\, .
\end{eqnarray}
With $\rho_{\rm inf}^{1/4} \lesssim 10^{16}\, \rm GeV$, which can be indicated from the current upper limit on tensor-to-scalar ratio in the single-field slow-roll models of inflation \cite{Planck:2018jri}, then we can derive the condition $M_{\rm PBH} \gtrsim 10 \, \rm g$.
Therefore, the PBH mass range we considered in this work is \cite{Papanikolaou:2020qtd}
\begin{eqnarray}
10 \, {\rm g} \lesssim M_{\rm PBH} \lesssim 10^9 \, \rm g \, .
\end{eqnarray}

\subsection{Schwarzschild black holes}

For the simplest black hole scenario Schwarzschild black holes, they are spherically symmetric and can only be described by the mass. 
The horizon temperature is defined by \cite{Hawking:1975vcx}
\begin{eqnarray}
T_{\rm PBH}= \frac{M_{\rm Pl}^2}{M_{\rm PBH}} \simeq 10^{13} \left(\frac{M_{\rm PBH}}{1 \, {\rm g}} \right)^{-1}  \rm GeV\, .
\end{eqnarray}
After formation, they emit the energetic particles through Hawking evaporation.
The emission rate of particles with type $i$ and spin $s$ from a black hole per units of energy $E$ and time $t$ is given by \cite{Hawking:1974rv}
\begin{eqnarray}
\frac{{\rm d} N_{i}}{{\rm d} E {\rm d}t}=\frac{1}{2\pi} \sum_{\rm dof}\frac{\Gamma_{s_i}(M_{\rm PBH},E)}{\exp\left({E/T_{\rm PBH}}\right)-(-1)^{2s_i}}\, ,
\label{dnde}
\end{eqnarray}
where the sum is over the particle total multiplicity, and $\Gamma_{s_i}(M_{\rm PBH},E)$ is the dimensionless absorption coefficient, called greybody factor
\begin{eqnarray}
\Gamma_{s_i}(M_{\rm PBH},E) = \frac{\sigma_{s_i}(M_{\rm PBH},E)E^2}{\pi} \, . 
\end{eqnarray} 
The greybody factor represents the spectral distortion of the blackbody, corresponding to the probability that spherical wave of the  particles induced by the thermal fluctuation of the vacuum at the black hole horizon escape the gravitational well.
The absorption cross-section $\sigma_{s_i}(M_{\rm PBH},E)$ would approach to the geometric optics limit $\sigma_g=27\pi G^2 M_{\rm PBH}^2$ in the high energy $E\gg T_{\rm PBH}$.
While it is a complicated function of $E$, $M_{\rm PBH}$, and spin $s$ in the low energy region.
 
With eq.~(\ref{dnde}), we can derive the mass loss rate of an evaporating PBH through Hawking evaporation \cite{Page:1976ki}
\begin{eqnarray}
\frac{{\rm d}M_{\rm PBH}}{{\rm d}t}= -\frac{f(M_{\rm PBH})}{M_{\rm PBH}^2} \, ,
\end{eqnarray} 
where $f(M_{\rm PBH})$ is the Page factor \cite{Page:1976ki, MacGibbon:1990zk}
\begin{eqnarray}
\begin{aligned}
f(M_{\rm PBH})&=-M_{\rm PBH}^2\frac{{\rm d}M_{\rm PBH}}{{\rm d}t}\\&=M_{\rm PBH}^2 \int_0^{+\infty}\frac{E}{2\pi}\sum_i\sum_{\rm dof}\frac{\Gamma_{s_i}(M_{\rm PBH},E)}{\exp\left({E/T_{\rm PBH}}\right)-(-1)^{2s_i}}{\rm d}E  \, ,
\end{aligned}
\end{eqnarray} 
which represents the number of relativistic degrees of freedom that can be emitted in the evaporation process as a function of the instantaneous black hole mass.

\begin{figure*}[!htbp]
\centering
\subfigcapskip=0pt
\subfigbottomskip=0pt
\subfigure[Schwarzschild black holes.]{\includegraphics[width=7.5cm]{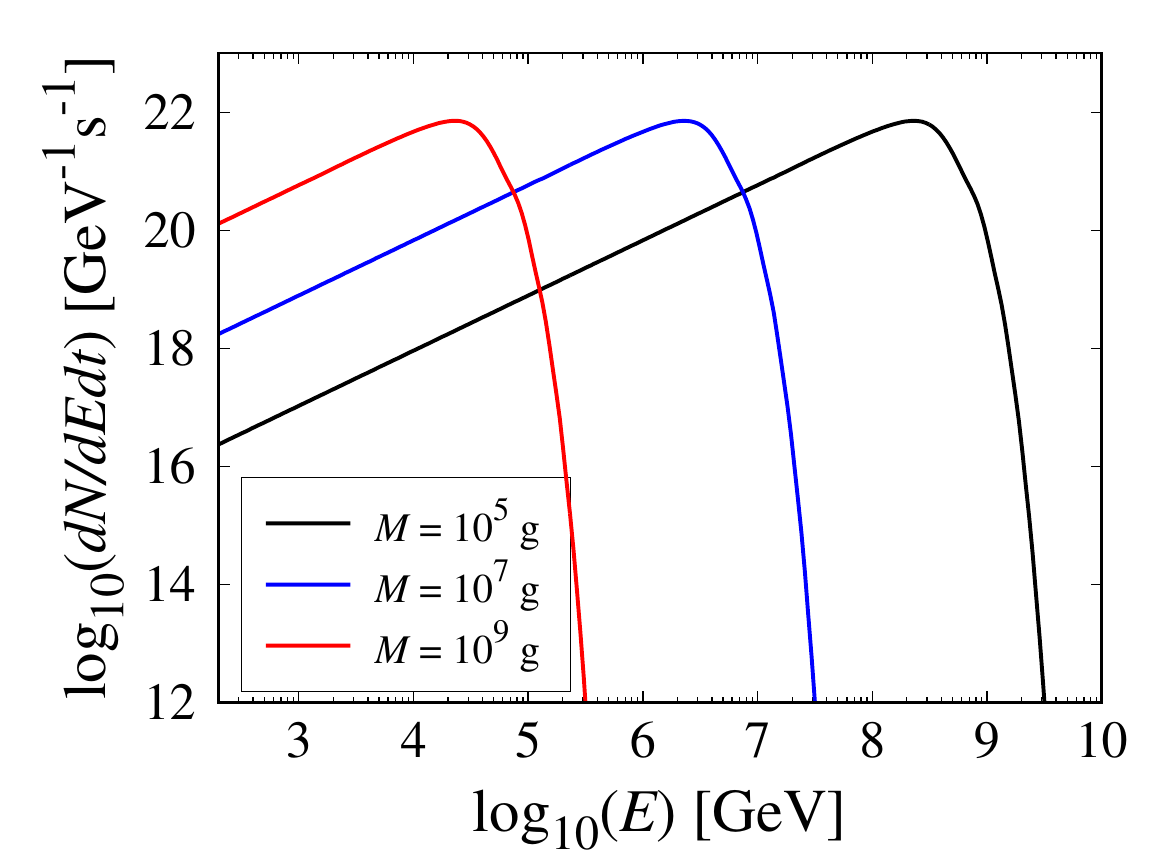}}\subfigure[Kerr black holes with $a_*=0.50$.]{\includegraphics[width=7.5cm]{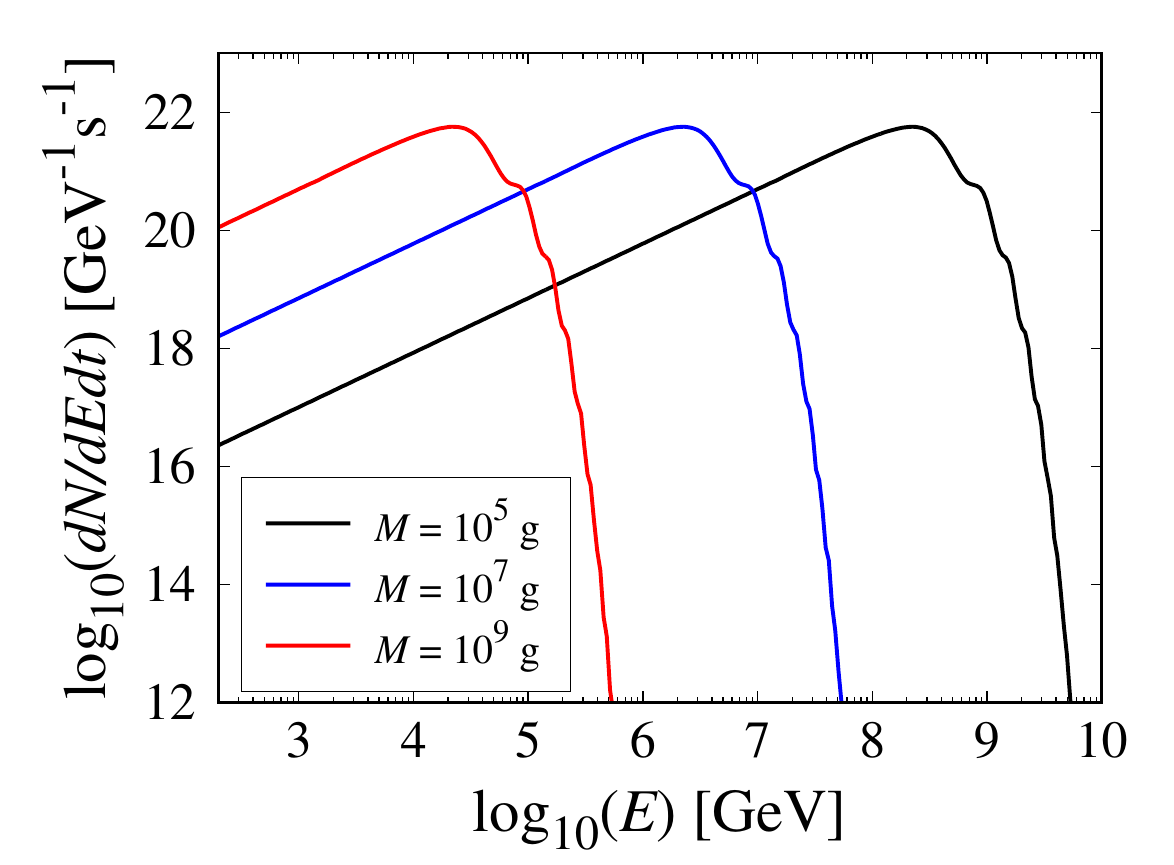}}
\subfigure[Kerr black holes with $a_*=0.75$.]{\includegraphics[width=7.5cm]{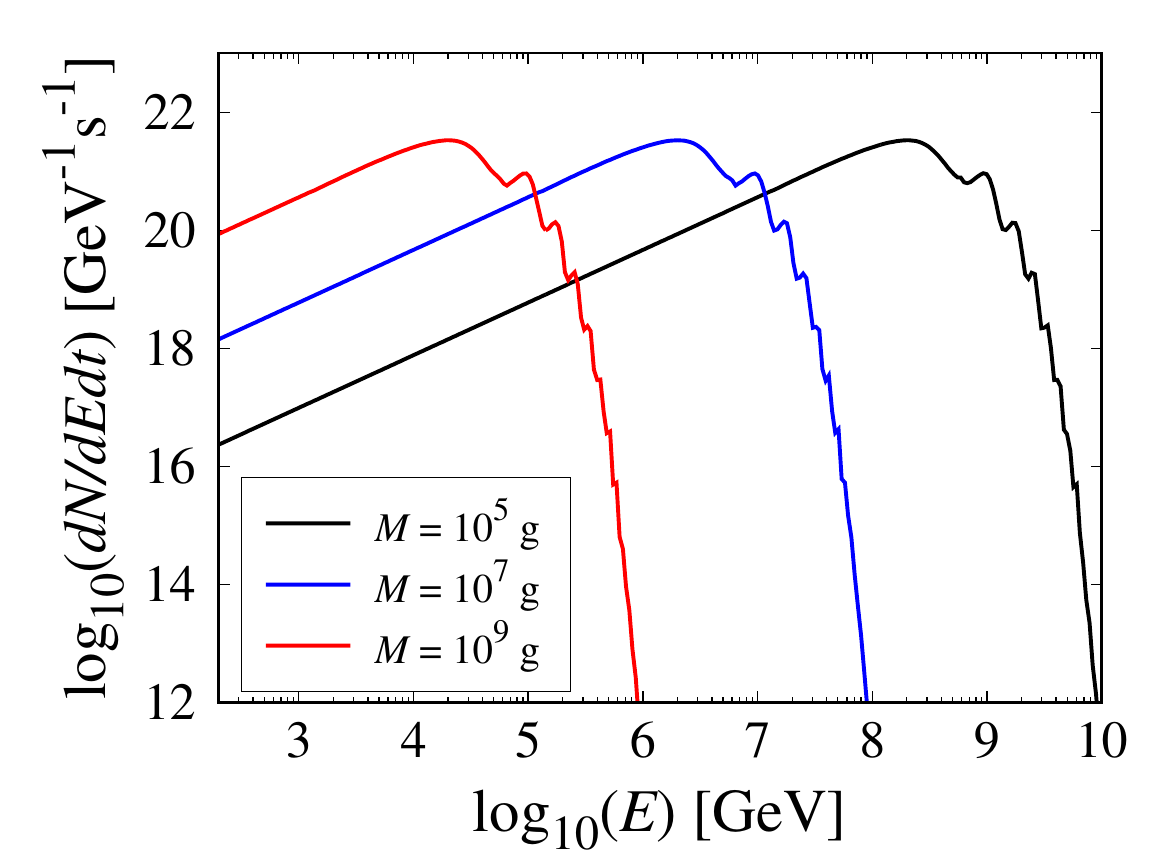}}\subfigure[Kerr black holes with $a_*=0.99$.]{\includegraphics[width=7.5cm]{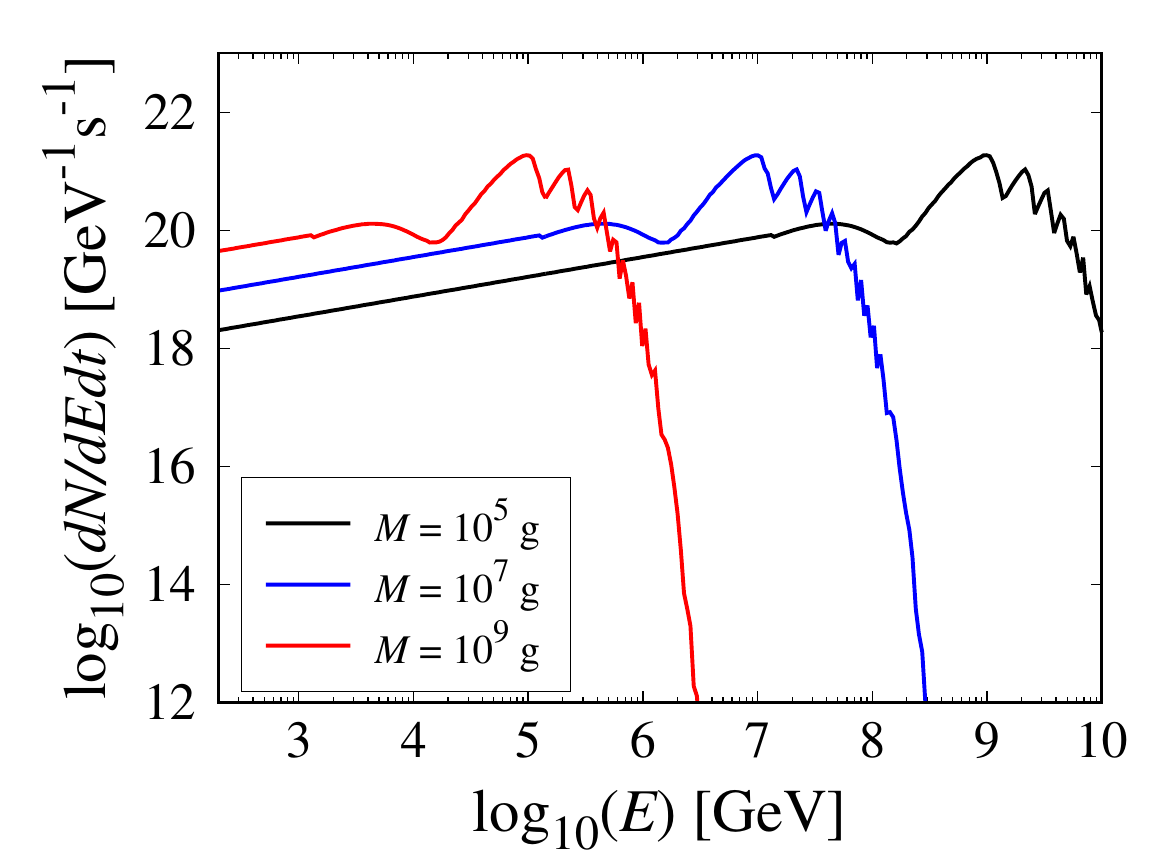}}
  \caption{The instantaneous spectra of the spin-0 particle emitted by a black hole as a function of energy. The different panels correspond to the Schwarzschild black holes and Kerr black holes (with $a_*=0.5$, $0.75$, and $0.99$). The black, blue, and red lines correspond to the PBH masses $M_{\rm PBH}=10^5\, \rm g$, $10^7\, \rm g$, and $10^9\, \rm g$, respectively.}
  \label{fig_spe}
\end{figure*}

We show the instantaneous spectra of spin-0 particle emitted by a Schwarzschild black hole as a function of energy in figure~\ref{fig_spe} with the public code {\tt BlackHawk} \cite{Arbey:2019mbc, Arbey:2021mbl}.
Three typical values of PBH mass are selected for comparisons.
The black, blue, and red lines represent the Schwarzschild PBH masses $M_{\rm PBH}=10^5\, \rm g$, $10^7\, \rm g$, and $10^9\, \rm g$, respectively.

\subsection{Kerr black holes}

Kerr black holes is an extension of the Schwarzschild black holes with the non-zero angular momentum, which could acquire the spin through accretion \cite{Page:1974he} or mergers \cite{Buonanno:2007sv}.
Here the horizon temperature can be described by
\begin{eqnarray}
T_{\rm PBH}= \frac{1}{4\pi M_{\rm PBH}}\frac{\sqrt{1-a_*}}{1+\sqrt{1-a_*}}\, ,
\end{eqnarray}
where $a_*$ is the reduced black hole spin parameter (the spin $a\equiv J/M_{\rm PBH}\in \left[0,M_{\rm PBH} \right]$)
\begin{eqnarray}
a_*\equiv\frac{J}{M_{\rm PBH}^2} \in \left[0,1 \right] \, ,
\end{eqnarray}
with the total angular momentum of the black hole $J$.
Then the emission rate of the particles becomes
\begin{eqnarray}
\frac{{\rm d} N_{i}}{{\rm d} E {\rm d}t}=\frac{1}{2\pi} \sum_{\rm dof}\frac{\Gamma_{s_i}^{lm}(M_{\rm PBH},E,a_*)}{\exp\left({E'/T_{\rm PBH}}\right)-(-1)^{2s_i}}\, ,
\label{dnde2}
\end{eqnarray}
where
\begin{eqnarray}
E' = E-\frac{m}{2M_{\rm PBH}}\frac{a_*}{1+\sqrt{1-a_*^2}} \, ,
\end{eqnarray}
with the particles total $l$ and axial $m$ angular momentum numbers, $m\in \{-l, ... , l\}$.

Using eq.~(\ref{dnde2}), the mass loss rate of these PBHs can be described by
\begin{eqnarray}
\frac{{\rm d}M_{\rm PBH}}{{\rm d}t}= -\frac{f(M_{\rm PBH}, a_*)}{M_{\rm PBH}^2} \, ,
\end{eqnarray} 
with the Page factor \cite{Page:1976ki}
\begin{eqnarray}
\begin{aligned}
f(M_{\rm PBH},a_*)&=-M_{\rm PBH}^2\frac{{\rm d}M_{\rm PBH}}{{\rm d}t}\\&=M_{\rm PBH}^2 \int_0^{+\infty}\frac{E}{2\pi}\sum_i\sum_{\rm dof}\frac{\Gamma_{s_i}^{lm}(M_{\rm PBH},E,a_*)}{\exp\left({E'/T_{\rm PBH}}\right)-(-1)^{2s_i}}{\rm d}E  \, .
\end{aligned}
\end{eqnarray} 
The evolution of $a_*$ is given by \cite{Page:1976df}
\begin{eqnarray}
\begin{aligned}
\frac{{\rm d} a_*}{{\rm d}t} &= \frac{{\rm d} (J/M_{\rm PBH}^2)}{{\rm d}t} = \frac{1}{M_{\rm PBH}^2}\frac{{\rm d} J}{{\rm d}t}-2\frac{J}{M_{\rm PBH}^3}\frac{{\rm d} M_{\rm PBH}}{{\rm d}t} \\
&=a_*\frac{2f(M_{\rm PBH},a_*)-g(M_{\rm PBH},a_*)}{M_{\rm PBH}^3} \, ,
\end{aligned}
\end{eqnarray} 
with
\begin{eqnarray}
\begin{aligned}
g(M_{\rm PBH},a_*) &=-\frac{M_{\rm PBH}}{a_*}\frac{{\rm d} J}{{\rm d}t}\\&=-\frac{M_{\rm PBH}}{a_*}\int_0^{+\infty}\frac{m}{2\pi} \sum_i \sum_{\rm dof}\frac{\Gamma_{s_i}^{lm}(M_{\rm PBH},E,a_*)}{\exp\left({E'/T_{\rm PBH}}\right)-(-1)^{2s_i}}{\rm d}E \, .
\end{aligned}
\end{eqnarray} 

For comparison, we also show the instantaneous spectra of spin-0 particle emitted by a Kerr black hole in figure~\ref{fig_spe} with the black hole spins $a_*=0.50$, $0.75$, and $0.99$. 
In these panels, the black, blue, and red lines represent the Kerr PBH masses $M_{\rm PBH}=10^5\, \rm g$, $10^7\, \rm g$, and $10^9\, \rm g$, respectively.
As we can see, the spectral distributions in the high energy regions would shift to the right for the black holes with large spin values, which is caused by the competition between the power-law behaviour of the greybody factor and the exponential cut-off of the Boltzmann factor. 
The Boltzmann factor is not greatly affected by the black hole spin, while the greybody factor is enhanced as the spin increases.
Note that the nonphysical ``oscillations" of the instantaneous spectra are the consequences of the sum over many angular modes for the Hawking radiation.
Since each one of these angular modes corresponds to one ``peak" in the spectrum, the sum over all angular modes would generate a fake oscillatory distribution. 
Therefore, the apparent ``oscillations" have the nonphysical significance, even if they increase with the black hole spin.

\subsection{ALPs emitted by PBHs}
\label{sec_ALP_from_PBH}
    
In this subsection, we briefly introduce the ALPs emitted by PBHs, see ref.~\cite{Schiavone:2021imu} for more details.
The ALP number density obeys the Boltzmann equation
\begin{eqnarray}
\left[\frac{\partial}{\partial t}-H(t)k\frac{\partial}{\partial k}+2H(t) \right]\frac{{\rm d}n_{\rm ALP}(k,t)}{{\rm d}k} = n_{\rm PBH}(t)\phi_0[k,M_{\rm PBH}(t)]\, ,
\label{alpbe}
\end{eqnarray}
with
\begin{eqnarray}
n_{\rm PBH}(t)=\frac{n_{\rm PBH}(t_{\rm in})}{a^3(t)}\, ,
\end{eqnarray}
where $\phi_0$ is the emission rate of ALP, $k$ is the modulus of ALP momentum, $n_{\rm PBH}$ and $n_{\rm ALP}$ are the number densities of PBH and ALP, respectively. 
The solution of eq.~(\ref{alpbe}) is
\begin{eqnarray}
\frac{{\rm d}n_{\rm ALP}(k',t_{\rm ev})}{{\rm d}k} = n_{\rm PBH}(t_{\rm ev}) \int_{t_{\rm in}}^{t_{\rm ev}} \frac{a'}{a(t)}\phi_0[k' \frac{a'}{a(t)},M_{\rm PBH}(t)] {\rm d}t \, ,
\label{alpbes}
\end{eqnarray}
where $a'$ corresponds to the PBH evaporation time $t_{\rm ev}$.
Considering the very light ALP, we have $k \simeq E$ with the ALP energy $E$.
The ALP energy density per unit of energy can be described by
\begin{eqnarray}
\frac{{\rm d}\rho_{\rm ALP}}{{\rm d}E} = E\frac{{\rm d}n_{\rm ALP}(E,t)}{{\rm d}k} \, .
\end{eqnarray}

Here we suppose the minimal scenario that ALPs could interact only with photons. 
Considering the ALP to photon oscillations in the magnetic field, the photon flux then can be described by 
\begin{eqnarray}
\frac{{\rm d}n_\gamma(E,t)}{{\rm d}E}=P_{a\gamma} \frac{{\rm d}n_{\rm ALP}(E,t)}{{\rm d}E}\, ,
\end{eqnarray}
where $P_{a\gamma}$ is the ALP-photon oscillation probability.
The detailed ALP-photon oscillations in the cosmic magnetic field can be found in section~\ref{sec_oscillations}.

\section{Primordial magnetic field}
\label{sec_PMF} 
 
In this section, we describe the PMF model and the latest upper limits on PMF, which is intended to discuss the ALP-photon oscillations in the PMF in next section. 
One simplest PMF is the homogeneous magnetic field model \cite{Barrow:1997mj, Kahniashvili:2008sh}, which can not be included in the homogeneous and isotropic expanding Universe filled with the cosmic plasma.
While the most widely accepted PMF model is the stochastic background field model, which can be modelled as the completely non-homogeneous component of the cosmic plasma \cite{Lewis:2004ef, Paoletti:2008ck, Planck:2015zrl}. 
In this case, the contributions of energy density and anisotropic stress at the homogeneous level can be neglected, the energy momentum tensor (EMT) components are on the same basis as the cosmic perturbations.

\subsection{Magnetic field model}

The magnetic field fluctuation for the stochastic background field with the non-helical and helical components can be described by the two-point correlation functions in the Fourier space \cite{Paoletti:2008ck, Planck:2015zrl}
\begin{eqnarray}
\langle B_i(\textbf{k})B^*_j(\textbf{k}') \rangle_1 = \frac{\left(2\pi \right)^3}{2}\delta^{(3)}(\textbf{k}-\textbf{k}')P_{ij}P_B(k) \, ,
\end{eqnarray}
\begin{eqnarray}
\langle B_i(\textbf{k})B^*_j(\textbf{k}') \rangle_2 = \frac{\left(2\pi \right)^3}{2}\delta^{(3)}(\textbf{k}-\textbf{k}')\left[P_{ij}P_B(k)+i\epsilon_{ijl}\hat{k}_lP_H(k)\right] \, ,
\end{eqnarray}
where $P_{ij}=\delta_{ij}-\hat{k}_i\hat{k}_j$ is the projection operator, $\hat{k}_i$ is the cartesian component of the normalized wave vector, $\epsilon_{ijl}$ is the total anti-symmetric tensor, $P_B$ and $P_H$ are the non-helical and helical components of the magnetic field power spectra
\begin{eqnarray}
P_B(k)=A_B k^{n_B} \, , 
\end{eqnarray}
\begin{eqnarray}
P_H(k)=A_H k^{n_H} \, ,
\end{eqnarray}
with the amplitudes $A_B$ and $A_H$, and the spectral indexes $n_B$ and $n_H$. 

For the non-helical component of PMF with the cut-off scale $k_D$, the magnetic field fluctuation on the comoving scale of $\lambda=1\, \rm Mpc$ can be described by
\begin{eqnarray}
B_{\lambda}^2=\frac{1}{2\pi^2}\int_0^{\infty}k^2{\rm e}^{-k^2\lambda^2}P_B(k) {\rm d}k=\frac{A_B}{4\pi^2\lambda^{n_B+3}}\Gamma\left(\frac{n_B+3}{2}\right)\, .
\end{eqnarray}
The cut-off scale is \cite{Mack:2001gc}
\begin{eqnarray}
k_D=\left(5.5\times 10^4\right)^\frac{1}{n_B+5}\left(\frac{B_\lambda}{\rm nG}\right)^{-\frac{2}{n_B+5}}\left(\frac{2\pi}{\lambda /\rm Mpc}\right)^\frac{n_B+3}{n_B+5}h^{\frac{1}{n_B+5}}
\left(\frac{\Omega_bh^2}{0.022}\right)^\frac{1}{n_B+5} \rm Mpc^{-1}\, ,~~
\end{eqnarray}
with the reduced Hubble constant $h$, and the baryon density parameter $\Omega_b$.
The magnetic field fluctuation of the helical component on the comoving scale of $\lambda$ is given by
\begin{eqnarray}
B_{\lambda}^2=\frac{\lambda}{2\pi^2}\int_0^{\infty}k^3{\rm e}^{-k^2\lambda^2}|P_H(k)| {\rm d}k=\frac{|A_H|}{4\pi^2\lambda^{n_H+3}}\Gamma\left(\frac{n_H+4}{2}\right)\, ,
\end{eqnarray}
with the helical spectral index $n_H > -4$.

The magnetic EMT can be regarded as the description of perturbations that carry the energy density and anisotropic stress, and generate the Lorentz force on the charged particles.
For the non-helical component of PMF, the relevant Fourier components of the EMT can be described by \cite{Planck:2015zrl}
\begin{eqnarray}
\begin{aligned}
\Big|\rho_B(k)\Big|^2&=\frac{1}{1024\pi^5}\int P_B(p)P_B(|\textbf{k}-\textbf{p}|)\left(1+\mu^2\right){\rm d}^3p\, ,
\end{aligned}
\end{eqnarray}
\begin{eqnarray}
\begin{aligned}
\Big|L_B^{(\rm scalar)}(k)\Big|^2&=\frac{1}{128\pi^2a^8}\int P_B(p)P_B(|\textbf{k}-\textbf{p}|)\left[1+\mu^2+4\gamma\beta\left(\gamma\beta-\mu\right)\right]{\rm d}^3p \, ,
\end{aligned}
\end{eqnarray}
\begin{eqnarray}
\begin{aligned}
\Big|\Pi^{(\rm vector)}(k)\Big|^2&=\frac{1}{512\pi^5}\int P_B(p)P_B(|\textbf{k}-\textbf{p}|)\left[\left(1-\gamma^2\right)\left(1+\beta^2\right)-\gamma\beta\left(\gamma\beta-\mu\right)\right]{\rm d}^3p \, ,
\end{aligned}
\end{eqnarray}
\begin{eqnarray}
\Big|\Pi^{(\rm tensor)}(k)\Big|^2=\frac{1}{512\pi^5}\int P_B(p)P_B(|\textbf{k}-\textbf{p}|)\left(1+2\gamma^2+\gamma^2\beta^2\right){\rm d}^3p \, ,
\end{eqnarray}
with $\mu=\hat{\textbf{p}}\cdot(\textbf{k}-\textbf{p})/|\textbf{k}-\textbf{p}|$, $\beta=\hat{\textbf{k}}\cdot(\textbf{k}-\textbf{p})/|\textbf{k}-\textbf{p}|$, and $\gamma=\hat{\textbf{k}}\cdot\hat{\textbf{p}}$. 
While for the helical component of PMF, the relevant components of the EMT are
\begin{eqnarray}
\begin{aligned}
\Big|\rho_B(k)\Big|^2&=\frac{1}{1024\pi^5}\int P_B(p)P_B(|\textbf{k}-\textbf{p}|)\left(1+\mu^2\right){\rm d}^3p \\
&-\frac{1}{512\pi^5}\int \mu P_H(p)P_H(|\textbf{k}-\textbf{p}|){\rm d}^3p\, ,
\end{aligned}
\end{eqnarray}
\begin{eqnarray}
\begin{aligned}
\Big|L_B^{(\rm scalar)}(k)\Big|^2&=\frac{1}{128\pi^2a^8}\int P_B(p)P_B(|\textbf{k}-\textbf{p}|)\left[1+\mu^2+4\gamma\beta\left(\gamma\beta-\mu\right)\right]{\rm d}^3p \\
&+\frac{1}{64\pi^2a^8}\int p P_H(p)P_H(|\textbf{k}-\textbf{p}|)\left(\mu-2\gamma\beta\right){\rm d}^3p\, ,
\end{aligned}
\end{eqnarray}
\begin{eqnarray}
\begin{aligned}
\Big|\Pi^{(\rm vector)}(k)\Big|^2&=\frac{1}{512\pi^5}\int P_B(p)P_B(|\textbf{k}-\textbf{p}|)\left[\left(1-\gamma^2\right)\left(1+\beta^2\right)-\gamma\beta\left(\gamma\beta-\mu\right)\right]{\rm d}^3p ~~~\\
&+\frac{1}{512\pi^5}\int P_H(p)P_H(|\textbf{k}-\textbf{p}|)\left(\mu-\gamma\beta\right){\rm d}^3p\, ,
\end{aligned}
\end{eqnarray}
\begin{eqnarray}
\begin{aligned}
\Big|\Pi^{(\rm tensor)}(k)\Big|^2&=\frac{1}{512\pi^5}\int P_B(p)P_B(|\textbf{k}-\textbf{p}|)\left(1+2\gamma^2+\gamma^2\beta^2\right){\rm d}^3p \\
&+\frac{1}{128\pi^5}\int \gamma\beta P_H(p)P_H(|\textbf{k}-\textbf{p}|){\rm d}^3p\, .
\end{aligned}
\end{eqnarray}

\subsection{PMF limits}
\label{sec_PMF_limits} 

Here we list the latest PMF limits set by the Planck collaboration in 2015 and by Jedamzik \& Saveliev in 2019, which are defined as Planck~2015 and JS~2019, respectively.
\begin{itemize}
\item  Planck~2015 \cite{Planck:2015zrl}: The 95\% confidence level comoving field amplitude limits of the non-helical and helical components over the $1\, \rm Mpc$ region are $B_{1\, \rm Mpc} < 4.4\, \rm nG$ and $5.6\, \rm nG$, respectively. These results are come from the CMB angular power spectra analysis by using the Planck likelihood. 
\item  JS~2019 \cite{Jedamzik:2018itu}: The 95\% confidence level limits of the scale-invariant ($n=0$, where $n$ is the spectral index) and a violet Batchelor spectrum ($n=5$) PMFs are $B<47\, \rm pG$ and $8.9\, \rm pG$, respectively. These limits are derived from the CMB radiation based on the non-observation of the magnetic field induced clumping of baryonic density fluctuations during recombination epoch. 
\end{itemize}
Since the limits of JS~2019 are more than one order of magnitude more stringent than any prior PMF limits, we will use these results to discuss the ALP-photon oscillations in the PMF in next section.

\section{ALP-photon oscillations in primordial magnetic field}
\label{sec_oscillations}

In this section, we discuss the ALP-photon oscillations in the PMF.
We first describe the general process of ALP-photon oscillation in the magnetic field.
Then we show the numerical results of the oscillation probability.
Finally, we briefly discuss the other effects on ALP-photon oscillations.

\subsection{General ALP-photon oscillation}
\label{sec_general_oscillation} 

The ALP-photon interaction in the external magnetic field can be described by the following effective Lagrangian \cite{Raffelt:1987im}
\begin{eqnarray}
\mathscr{L}=-\frac{1}{4}F^{\mu\nu}F_{\mu\nu}+\frac{1}{2}\left(\partial_\mu a\partial^\mu a - m_a^2a^2\right) -\frac{1}{4}g_{a\gamma}aF_{\mu\nu}\tilde{F}^{\mu\nu}\, ,
\end{eqnarray}
with $-\frac{1}{4}g_{a\gamma}aF_{\mu\nu}\tilde{F}^{\mu\nu}=g_{a\gamma}a\textbf{E}\cdot\textbf{B}$, where $a$ is the ALP field, $m_a$ is the ALP mass, $g_{a\gamma}$ is the ALP-photon coupling constant, $F_{\mu\nu}$ is the electromagnetic field tensor and its dual tensor $\tilde{F}^{\mu\nu}=\frac{1}{2}\epsilon^{\mu\nu\rho\sigma}F_{\rho\sigma}$, and $\textbf{E}$, $\textbf{B}$ are the local electric and magnetic field vectors. 

Considering the ALP-photon system propagates along the direction $x_3$, the system $\psi(x_3)$ can be described by \cite{DeAngelis:2011id}
\begin{eqnarray}
\psi(x_3) = 
\left(
\begin{array}{c}
A_1(x_3)\\
A_2(x_3)\\
a(x_3)
\end{array}
\right)\, ,
\label{psix3}
\end{eqnarray}
where $A_1(x_3)$ and $A_2(x_3)$ correspond to the linear polarization amplitudes of the photons in the perpendicular directions ($x_1 \otimes x_2$)
\begin{eqnarray}
|A_1\rangle=
\left(
\begin{array}{c}
1 \\
0 \\
0  
\end{array}
\right)\, , \qquad
|A_2\rangle=
\left(
\begin{array}{c}
0 \\
1 \\
0  
\end{array}
\right)\, , \qquad
|a\rangle=
\left(
\begin{array}{c}
0 \\
0 \\
1  
\end{array}
\right)\, ,
\end{eqnarray}
and $a$ is the ALP.
The equation of motion (EOM) for the ALP-photon system in the magnetic field with the beam energy $E$ can be described by \cite{Raffelt:1987im}
\begin{eqnarray}
\left(i \frac{\rm d^2}{{\rm d} x_3^2}+E^2+2E\mathcal{M}(E,x_3)\right)\psi(x_3) =0\, .
\label{eom_alp-photon}
\end{eqnarray}
Considering the short-wavelength approximation, we have
\begin{eqnarray}
\begin{aligned}
\left(\frac{\rm d^2}{{\rm d} x_3^2}+E^2\right)\psi(x_3) &=\left(i \frac{\rm d}{{\rm d} x_3}+E\right)\left(-i \frac{\rm d}{{\rm d} x_3}+E\right)\psi(x_3)\\
&=2E\left(i \frac{\rm d}{{\rm d} x_3}+E\right)\psi(x_3)\, .
\end{aligned}
\end{eqnarray}
Then we can rewrite the second-order eq.~(\ref{eom_alp-photon}) to first-order as
\begin{eqnarray}
\left(i \frac{\rm d}{{\rm d} x_3}+E+\mathcal{M}(E,x_3)\right)\psi(x_3) =0\, ,
\end{eqnarray}
where $\mathcal{M}(E,x_3)$ is the mixing matrix 
\begin{eqnarray}
\mathcal{M}(E,x_3)=
\left(
\begin{array}{ccc}
\Delta_{11}(E,x_3)~  &  \Delta_{12}(E,x_3)~ & \Delta_{a \gamma, 1}(x_3) \\
\Delta_{21}(E,x_3)~  &  \Delta_{22}(E,x_3)~ & \Delta_{a \gamma, 2}(x_3) \\
\Delta_{a \gamma, 1}(x_3)~  &  \Delta_{a \gamma, 2}(x_3)~  & \Delta_{aa}(E)
\end{array}
\right)\, ,
\label{Mex3_1}
\end{eqnarray}
with the terms
\begin{eqnarray}
\Delta_{11}(E,x_3)&=&\Delta_{\rm pl}(E,x_3)+2\Delta_{\rm QED}(E,x_3)  + \Delta_{\rm CMB}(E)\, ,\\
\Delta_{22}(E,x_3)&=&\Delta_{\rm pl}(E,x_3)+\frac{7}{2} \Delta_{\rm QED}(E,x_3)  + \Delta_{\rm CMB}(E)\, ,\\
\Delta_{a \gamma, 1}(x_3)&=&\frac{1}{2}g_{a\gamma}B_{T,1}(x_3)=0\, ,\\
\Delta_{a \gamma, 2}(x_3)&=&\frac{1}{2}g_{a\gamma}B_{T,2}(x_3)=\frac{1}{2}g_{a\gamma}B_T(x_3)\, ,\\
\Delta_{aa}(E)&=&-\frac{m_a^2}{2E}\, ,
\end{eqnarray}
where $\Delta_{12}(E,x_3)$ and $\Delta_{21}(E,x_3)$ represent the Faraday rotation terms and can be neglected. 
Here we assume that the transversal magnetic field $B_T$ is aligned along $x_2$. 
We can rewrite the mixing matrix $\mathcal{M}(E,x_3)$ as
\begin{eqnarray}
\mathcal{M}(E,x_3)=
\left(
\begin{array}{ccc}
\Delta_{11}(E,x_3)~ &  0&0 \\
 0& \Delta_{22}(E,x_3)~ & \Delta_{a \gamma}(x_3) \\
 0&\Delta_{a\gamma}(x_3)  & \Delta_{aa}(E)
\end{array}
\right)\, ,
\label{Mex3_2}
\end{eqnarray}
with the terms
\begin{eqnarray}
\Delta_{\rm pl}(E,x_3)&=&-\frac{\omega_{\rm pl}^2(x_3)}{2E}\simeq -1.08 \times 10^{-1} \left(\frac{n_e}{\rm cm^{-3}}\right)\left(\frac{E}{\rm GeV}\right)^{-1}{\rm Mpc}^{-1}\, ,\\
\Delta_{a\gamma}(x_3)&=&\frac{1}{2}g_{a\gamma}B_T(x_3)\simeq 1.52 \times 10^{-2}  \left(\frac{g_{a\gamma}}{10^{-11}\, \rm{GeV}^{-1}}\right) \left(\frac{B_T(x_3)}{1\, \rm n G}\right){\rm Mpc}^{-1}\, ,\\
\Delta_{aa}(E)&=&-\frac{m_a^2}{2E}\simeq -0.78 \times 10^{2} \left(\frac{m_a}{10^{-9}\, \rm GeV}\right)^2\left(\frac{E}{\rm GeV}\right)^{-1}{\rm Mpc}^{-1}\, , \\
\Delta_{\rm QED}(E,x_3)&=&\frac{\alpha E}{45\pi} \left( \frac{B_T(x_3)}{B_{\rm cr}}\right)^2 \simeq4.10\times 10^{-12}  \left(\frac{E}{\rm GeV}\right)\left(\frac{B_T(x_3)}{1\, \rm n G}\right)^2{\rm Mpc}^{-1}\, ,\\
\Delta_{\rm CMB}(E)&=& \rho_{\rm CMB}E \simeq 0.80 \times 10^{-4}\left(\frac{E}{\rm GeV}\right) {\rm Mpc}^{-1}\, .
\end{eqnarray}

The term $\Delta_{\rm pl}(E,x_3)$ represents the plasma effect when ALP-photon system propagates in the plasma with the plasma frequency $\omega_{\rm pl}= \sqrt{4\pi \alpha n_e/m_e}$, where $\alpha$ is the fine-structure constant, $n_e$ is the free electron number density, and $m_e$ is the electron mass.
The term $\Delta_{\rm QED}(E,x_3)$ represents the QED vacuum polarization effect from the following Heisenberg-Euler-Weisskopf effective Lagrangian \cite{Schwinger:1951nm}
\begin{eqnarray} 
\mathscr{L}=\frac{2\alpha^2}{45m_e^2}\left[ \left(\textbf{E}^2-\textbf{B}^2\right)^2+7\left(\textbf{E} \cdot \textbf{B} \right)^2\right] \, ,
\end{eqnarray}
with the critical magnetic field $B_{\rm cr} = m^2_e/|e|\simeq4.41\times 10^{13} \rm \, G$.
The term $\Delta_{\rm CMB}(E,x_3)$ represents the CMB photon dispersion effect with \cite{Dobrynina:2014qba}
\begin{eqnarray} 
\rho_{\rm CMB}\simeq0.511\times10^{-42}\, .
\end{eqnarray}
 
Now considering the ALP-photon system propagates in a homogeneous magnetic field $B_T$, the ALP-photon oscillation probability can be simply described by
\begin{eqnarray}
\mathcal{P}_{a\gamma}(E,x_3)=\left( \frac{g_{a\gamma}B_T L_{\rm osc}(E)}{2\pi}\right)^2 \sin^2\left(\frac{\pi x_3}{L_{\rm osc}(E)}\right)\, ,
\end{eqnarray}  
with the oscillation length
\begin{eqnarray}
\begin{aligned}
L_{\rm osc}(E) &= 2\pi\left[\left(\Delta_{22}(E)-\Delta_{aa}(E)\right)^2+4\Delta_{a \gamma}^2\right]^{-\frac{1}{2}}\\
&= 2\pi\left[\left[ \frac{|m_a^2-\omega_{\rm pl}^2|}{2E}+E\left(\frac{7\alpha}{90\pi} \left( \frac{B_T}{B_{\rm cr}}\right)^2 +\rho_{\rm CMB}\right)\right]^2+g_{a\gamma}^2B_T^2\right]^{-\frac{1}{2}}  \, . ~
\label{losc}
\end{aligned}
\end{eqnarray}  

\subsection{Numerical results of ALP-photon oscillation probability}

Using the PMF limits JS~2019 given in section~\ref{sec_PMF_limits} and the ALP-photon oscillation process in the magnetic field discussed in section~\ref{sec_general_oscillation}, we can derive the oscillation probability for the ALP-photon system propagates in the PMF.
In this subsection, we show the numerical results of ALP-photon oscillation probability with the homogeneous and stochastic magnetic field scenarios, which is simulated with the domain-like structure.
   
In this case, the ALP-photon system in eq.~(\ref{psix3}) can be described by the following density matrix
\begin{eqnarray}
\rho(x_3) = 
\left(
\begin{array}{c}
A_1(x_3)\\
A_2(x_3)\\
a(x_3)
\end{array}
\right)\otimes
\left(
\begin{array}{c}
A_1(x_3), A_2(x_3), a(x_3)
\end{array}
\right)^* \, ,
\end{eqnarray} 
which satisfies the Von Neumann-like commutator equation
\begin{eqnarray}
i\frac{{\rm d}\rho(x_3)}{{\rm d}x_3}= \rho(x_3)\mathcal{M}^\dagger(E,x_3)-\mathcal{M}(E,x_3)\rho(x_3) \, .
\label{vne}
\end{eqnarray} 
The solution of eq.~(\ref{vne}) is
\begin{eqnarray}
\rho(x_3)=\mathcal{T}(E,x_3)\rho(0)\mathcal{T}^\dagger(E,x_3) \, ,
\end{eqnarray} 
with the initial density matrix $\rho(0)$
\begin{eqnarray}
\rho(0)=\frac{1}{2}
\left(
\begin{array}{ccc}
1~ & 0~ & 0 \\
0~ & 1~ & 0 \\
0~ & 0~ & 0 
\end{array}
\right)\, ,
\end{eqnarray}
and the transport matrix $\mathcal{T}(E,x_3)$
\begin{eqnarray}
\mathcal{T}(E,x_3)= \prod^{N}_{i=1}\mathcal{T}(E_i,x_{3,i}) \, .
\end{eqnarray} 
Then the ALP-photon oscillation probability $\mathcal{P}_{a\gamma}$ for some propagation distance can be described by
\begin{eqnarray}
\mathcal{P}_{a\gamma} = 1-{\rm Tr}\left[\left(\rho_{11}+\rho_{22}\right)\mathcal{T}(E,x_3)\rho(0)\mathcal{T}^\dagger(E,x_3) \right] \, ,
\end{eqnarray} 
with 
\begin{eqnarray}
{\rho}_{11} = 
\left(
\begin{array}{ccc}
1~ & 0~ & 0 \\
0~ & 0~ & 0 \\
0~ & 0~ & 0 
\end{array}
\right)\, , \qquad
{\rho}_{22} = 
\left(
\begin{array}{ccc}
0~ & 0~ & 0 \\
0~ & 1~ & 0 \\
0~ & 0~ & 0 
\end{array}
\right)\, .
\end{eqnarray}

\begin{figure*}[!htbp]
\centering
\subfigcapskip=0pt
\subfigbottomskip=0pt
\subfigure[{$E=10^{-3}\, \rm GeV$, $B_T=47\,\rm pG$.}]{\includegraphics[width=7.35cm]{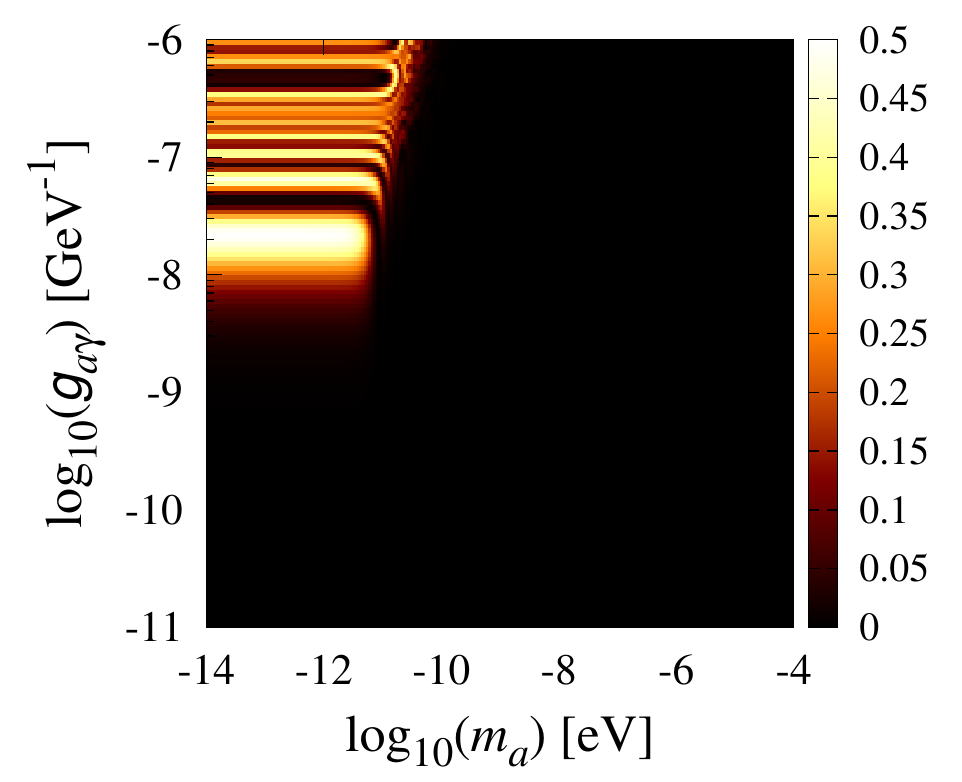}}\subfigure[{$E=10^{-3}\, \rm GeV$, $B_T=5\,\rm nG$.}]{\includegraphics[width=7.35cm]{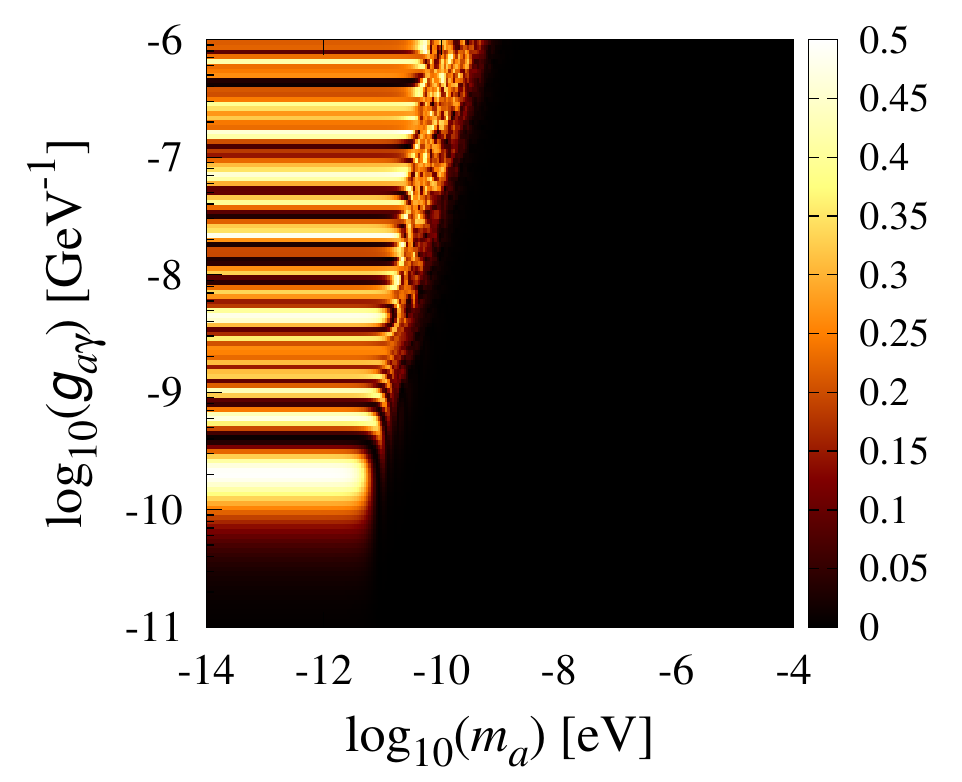}}
\subfigure[{$E=1\, \rm GeV$, $B_T=47\,\rm pG$.}]{\includegraphics[width=7.35cm]{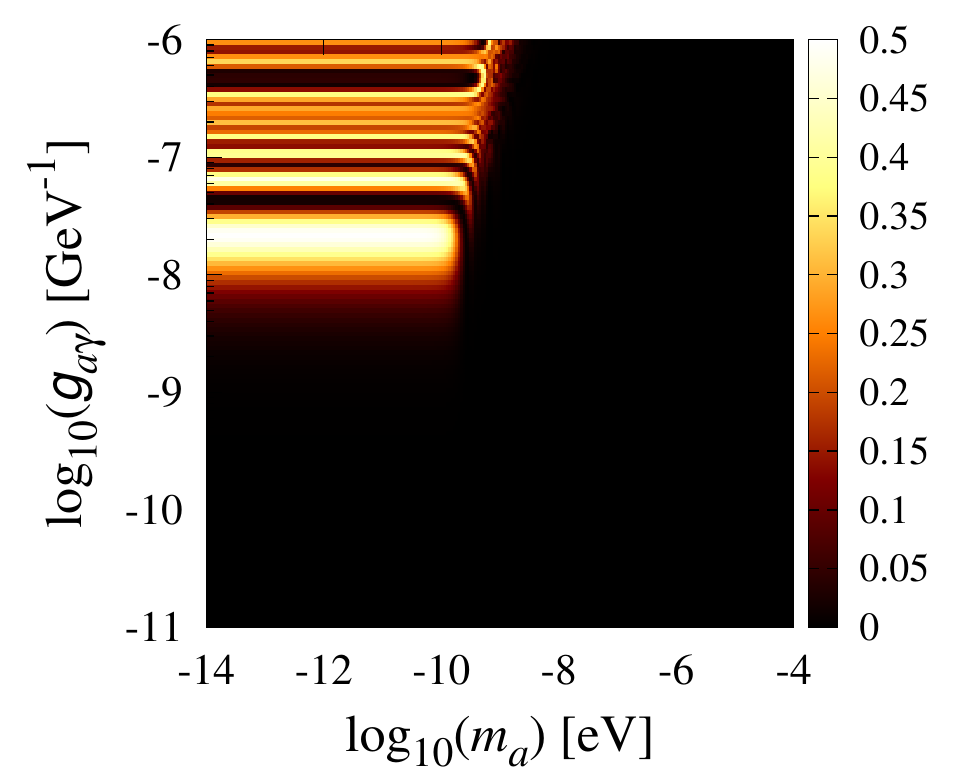}}\subfigure[{$E=1\, \rm GeV$, $B_T=5\,\rm nG$.}]{\includegraphics[width=7.35cm]{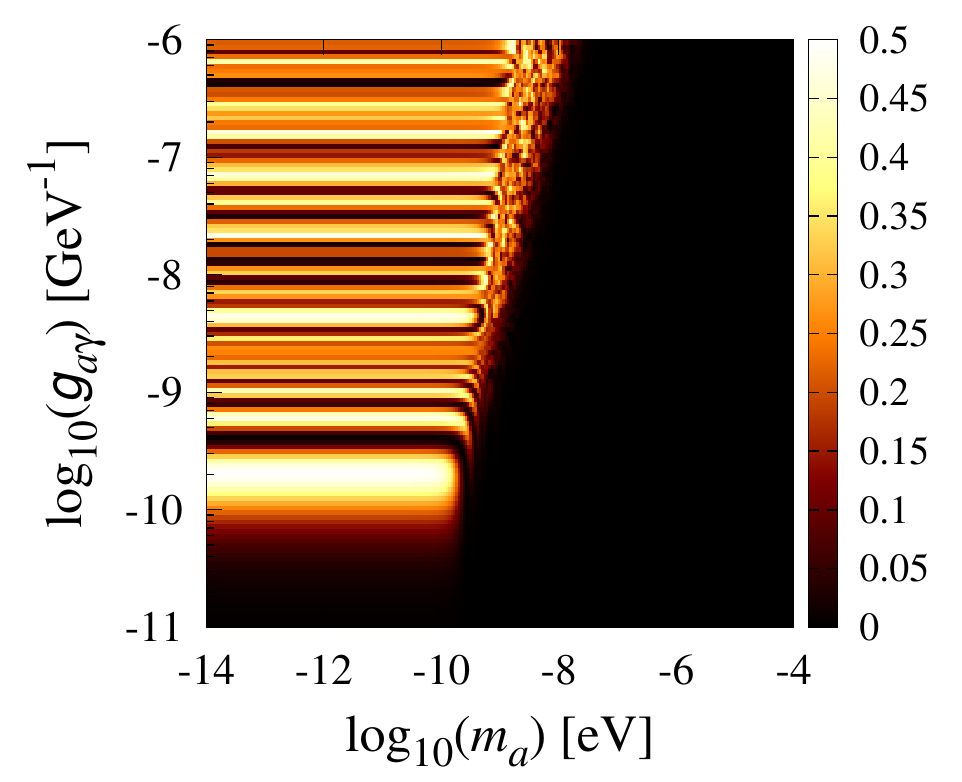}}
\subfigure[{$E=10^{3}\, \rm GeV$, $B_T=47\,\rm pG$.}]{\includegraphics[width=7.35cm]{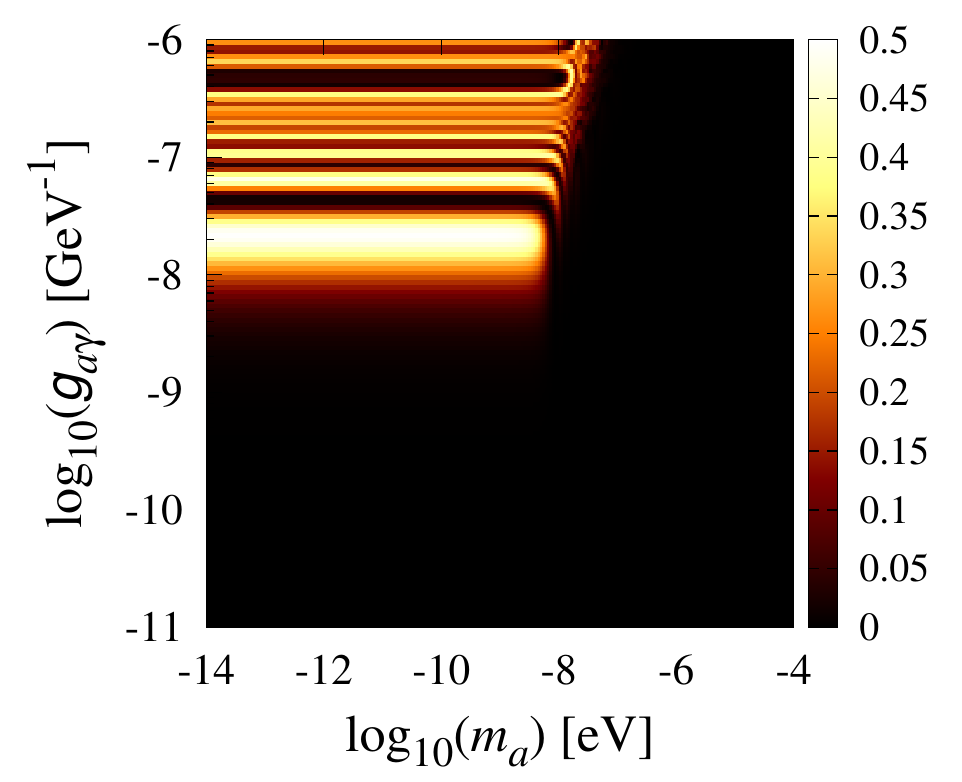}}\subfigure[{$E=10^{3}\, \rm GeV$, $B_T=5\,\rm nG$.}]{\includegraphics[width=7.35cm]{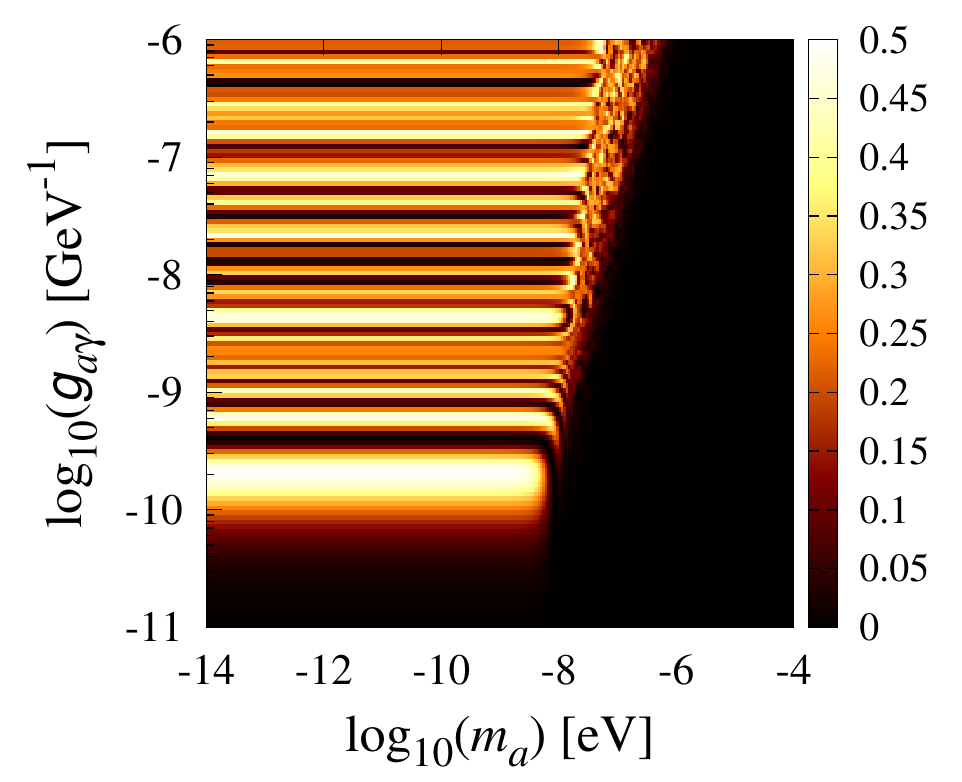}}
\caption{
The distributions of ALP-photon oscillation probability in the $m_a-g_{a\gamma}$ plane for several typical energy values. The top, middle, and bottom panels correspond to $E=10^{-3}\, \rm GeV$, $1\, \rm GeV$, and $10^{3}\, \rm GeV$, respectively. The left and right panels represent $B_T=47\,\rm pG$ and $5\,\rm nG$, respectively.
}
\label{fig_pa_contour_1}
\end{figure*}
   
\begin{figure*}[!htbp]
\centering
\subfigcapskip=0pt
\subfigbottomskip=0pt
\subfigure[{$E=10^{4}\, \rm GeV$, $B_T=47\,\rm pG$.}]{\includegraphics[width=7.35cm]{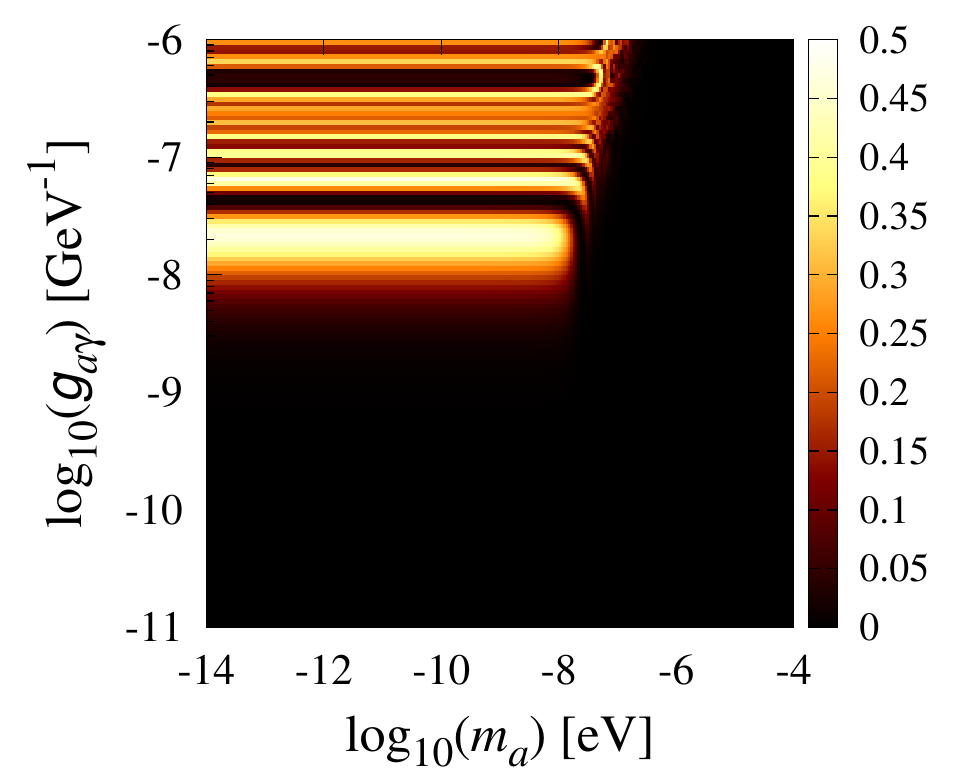}}\subfigure[{$E=10^{4}\, \rm GeV$, $B_T=5\,\rm nG$.}]{\includegraphics[width=7.35cm]{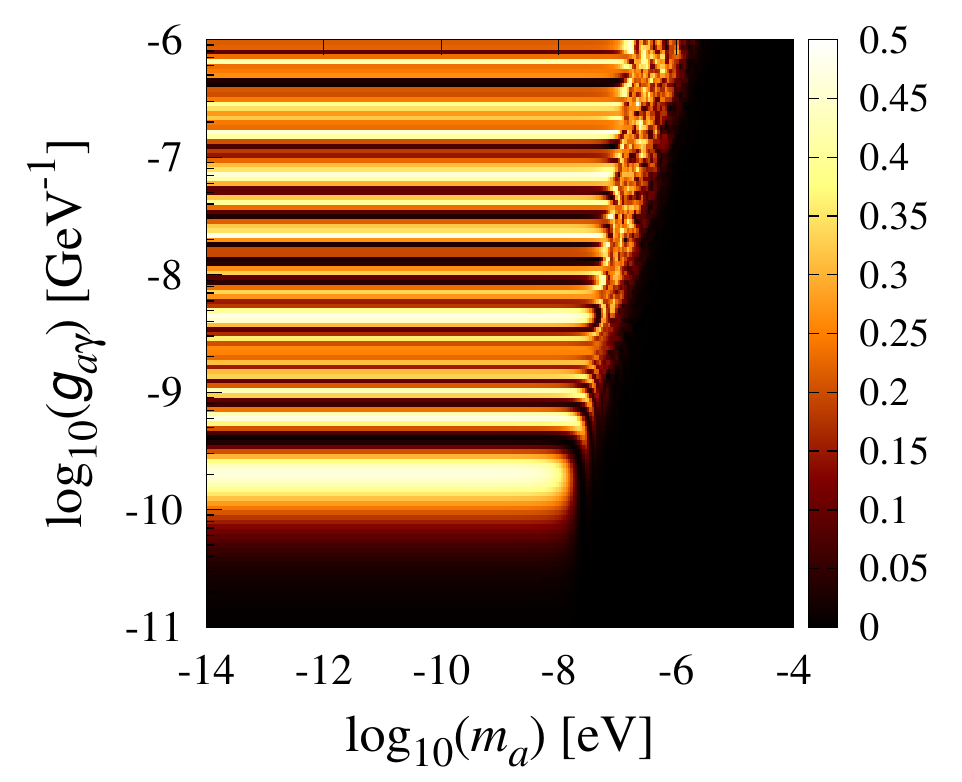}}
\subfigure[{$E=10^{5}\, \rm GeV$, $B_T=47\,\rm pG$.}]{\includegraphics[width=7.35cm]{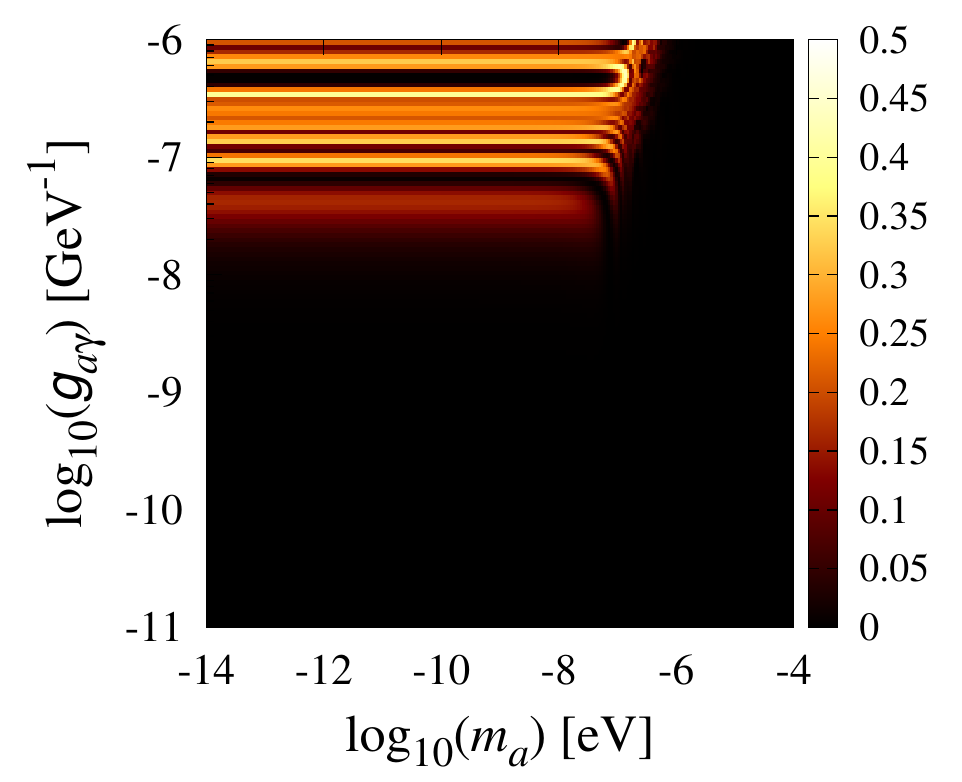}}\subfigure[{$E=10^{5}\, \rm GeV$, $B_T=5\,\rm nG$.}]{\includegraphics[width=7.35cm]{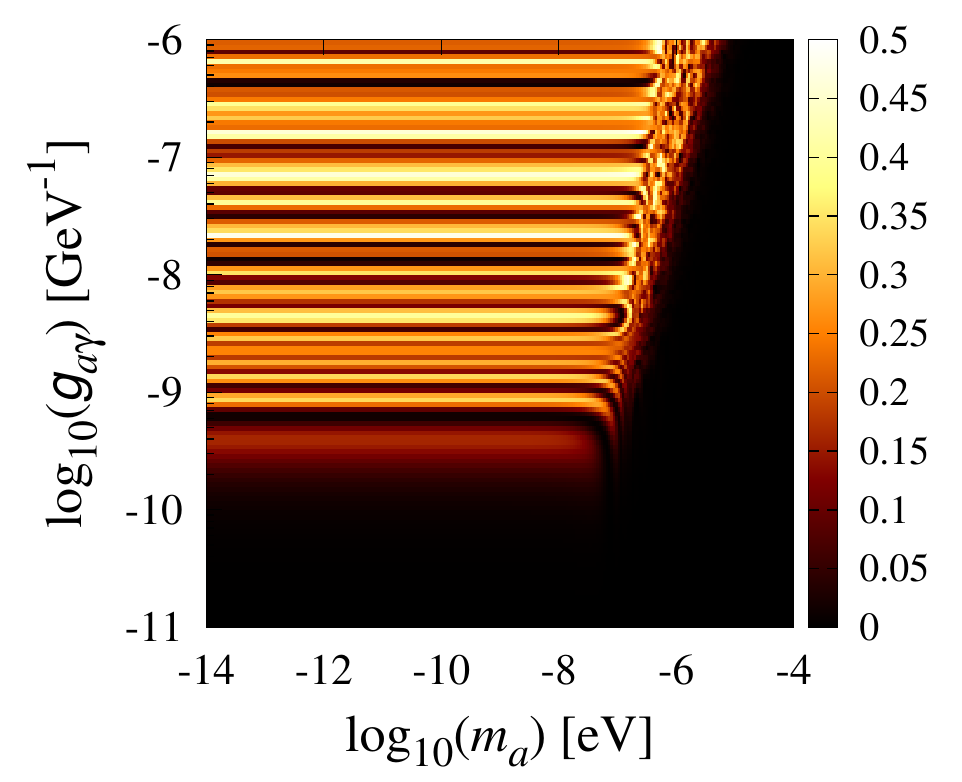}}
\subfigure[{$E=10^{6}\, \rm GeV$, $B_T=47\,\rm pG$.}]{\includegraphics[width=7.35cm]{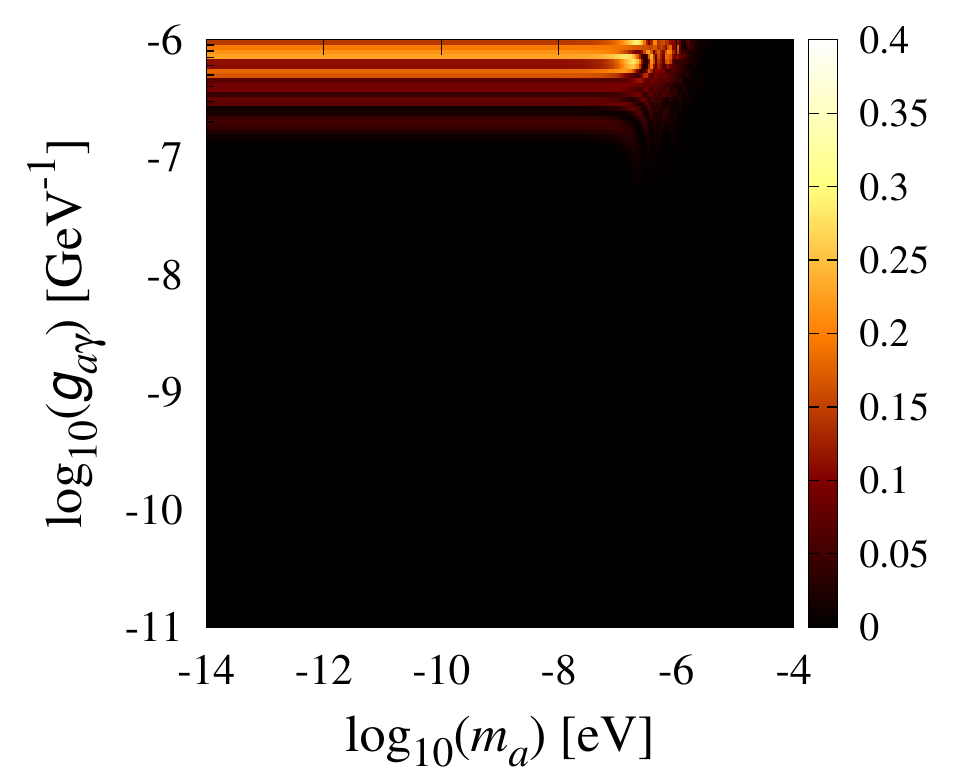}}\subfigure[{$E=10^{6}\, \rm GeV$, $B_T=5\,\rm nG$.}]{\includegraphics[width=7.35cm]{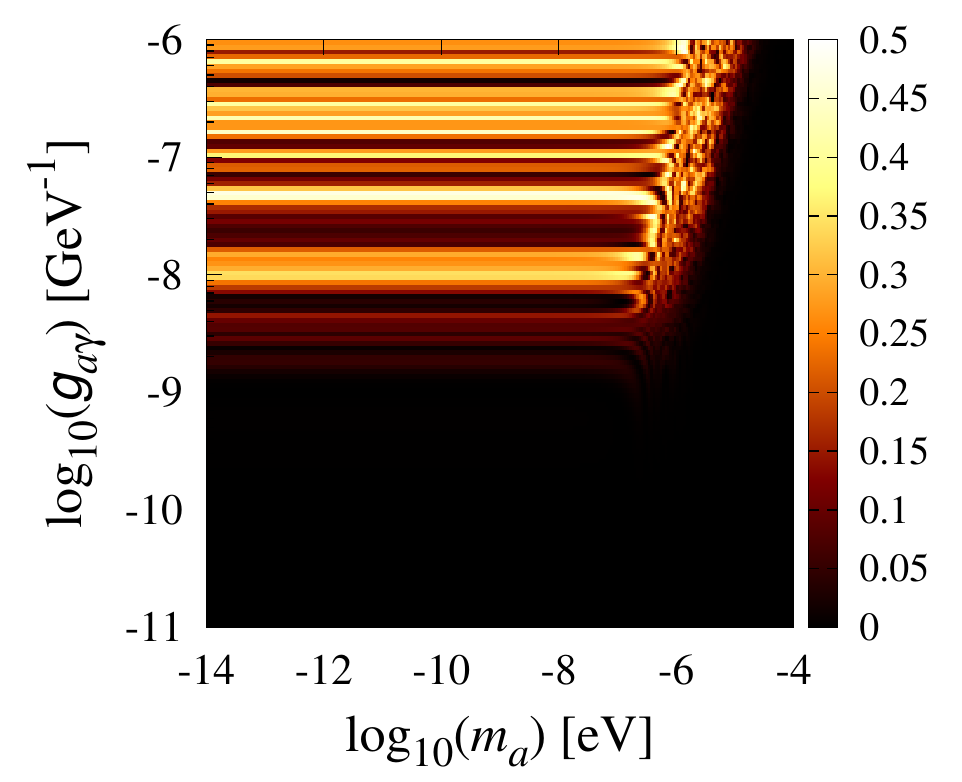}}
\caption{
Same as figure~\ref{fig_pa_contour_1} but for different energy values. The top, middle, and bottom panels correspond to $E=10^{4}\, \rm GeV$, $10^{5}\, \rm GeV$, and $10^{6}\, \rm GeV$, respectively. The left and right panels represent $B_T=47\,\rm pG$ and $5\,\rm nG$, respectively.
}
\label{fig_pa_contour_2}
\end{figure*}
         
Here we show the numerical results of the ALP-photon oscillation probability for the homogeneous magnetic field with the PMF limits JS~2019.
In this case, the magnetic field strength and direction in each domain are same, which corresponds to the scale-invariant limit $B<47\, \rm pG$ with the spectral index $n=0$.
In figure~\ref{fig_pa_contour_1}, we plot the oscillation probability distributions in the ALP parameter ($m_a$, $g_{a\gamma}$) space for several typical values of energy.
The top, middle, and bottom panels correspond to $E=10^{-3}\, \rm GeV$, $1\, \rm GeV$, and $10^{3}\, \rm GeV$, respectively.
For comparison, the oscillation probability distributions with the PMF limits Planck~2015 ($B\sim5\, \rm nG$) are also shown.
The left and right panels correspond to the magnetic field strength $B_T=47\,\rm pG$ and $B_T=5\,\rm nG$, respectively. 
The other typical values of energy ($E=10^{4}\, \rm GeV$, $10^{5}\, \rm GeV$, and $10^{6}\, \rm GeV$) for the oscillation probability distributions can be found in figure~\ref{fig_pa_contour_2}.

In these panels, the distributions of ALP-photon oscillation probability at $0.01 \lesssim \mathcal{P}_{a\gamma} \lesssim 0.50$ in the $m_a-g_{a\gamma}$ plane are shown, while the probability at $\mathcal{P}_{a\gamma} \lesssim 0.01$ could not be completely displayed.
The oscillation probability distribution changes significantly with the energy, and could be dramatically suppressed in the energy above $\sim10^4\, \rm GeV$. 
We also find the significant difference between the strength $B_T=47\,\rm pG$ and $5\,\rm nG$ with the homogeneous magnetic field scenario.
Here we do not show the specific oscillation probability, which can be found in next subsection with the stochastic magnetic field scenario.

\begin{figure*}[t]
\centering
\includegraphics[width=12.8cm]{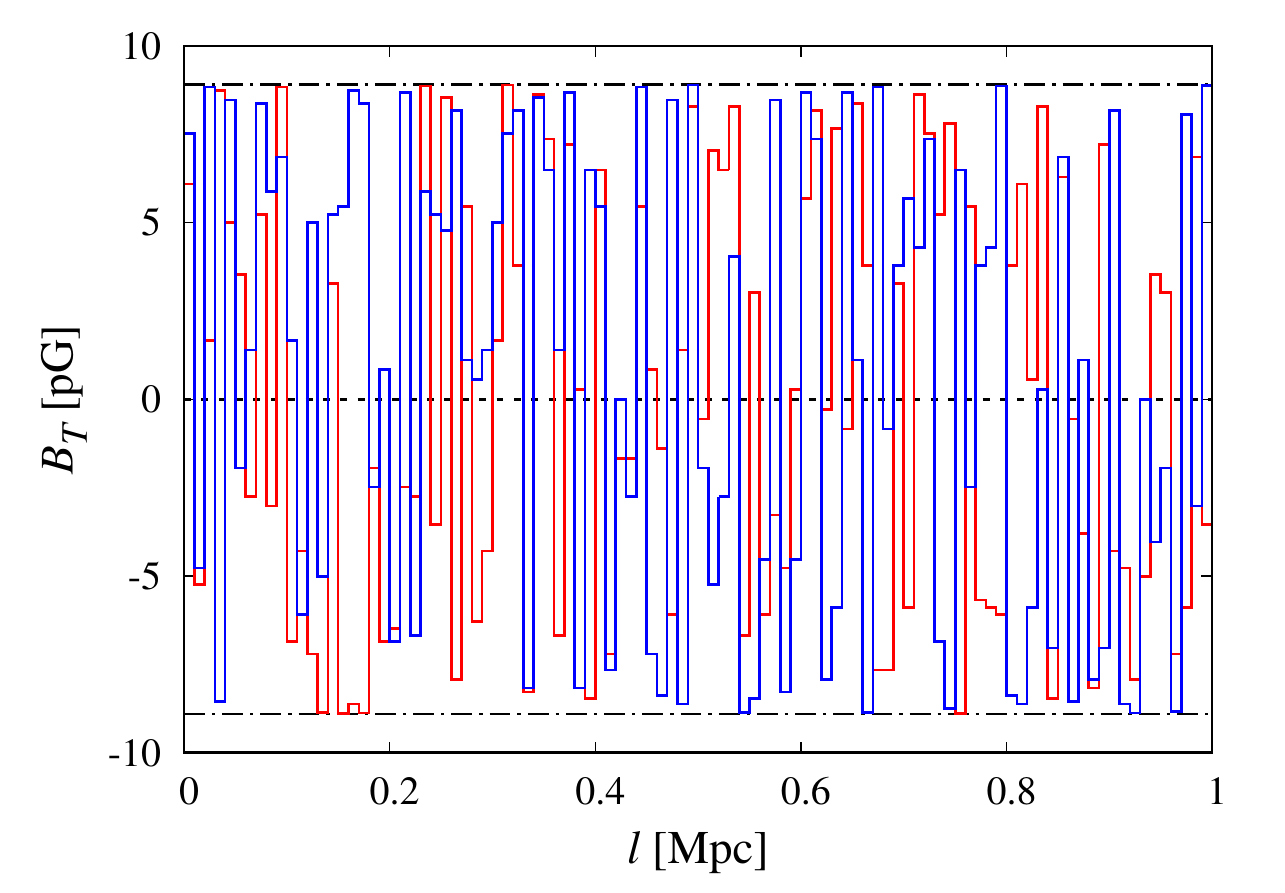}
\caption{
Effective magnetic field strength distributions for the stochastic magnetic field scenario with 100 domains for $1\, \rm Mpc$. The blue and red lines correspond to the magnetic field strength in each domain with the two stochastic magnetic field configurations. The dot-dashed black lines represent $|B_T| = 8.9\, \rm pG$.}
\label{fig_b_ran}
\end{figure*}

\subsubsection{Stochastic magnetic field}

For the stochastic magnetic field scenario, if the transversal magnetic field $B_T$ is not aligned along the direction of $x_2$ and forms the angle $\phi$, the mixing matrix $\mathcal{M}(E,x_3)$ in eq.~(\ref{Mex3_1}) should be converted as
\begin{eqnarray}
\mathcal{M}(E,x_3)=\mathcal{V}^\dag(\phi)\mathcal{M}'(E,x_3)\mathcal{V}(\phi)\, ,
\end{eqnarray}
with
\begin{eqnarray}
\mathcal{V}(\phi)=
\left(
\begin{array}{ccc}
\cos \phi~ &  -\sin \phi~  & 0  \\
\sin \phi~ &  \cos \phi~  & 0  \\
0 &  0  & 1  \\
\end{array}
\right)\, .
\end{eqnarray}
Then we can rewrite the mixing matrix $\mathcal{M}(E,x_3,\phi)$ in eq.~(\ref{Mex3_2}) as
\begin{eqnarray}
\mathcal{M}(E,x_3,\phi)=
\left(
\begin{array}{ccc}
\Delta_{11}(E,x_3)  &  0 & \Delta_{a \gamma}(x_3)\sin \phi \\
0  &  \Delta_{22}(E,x_3) & \Delta_{a \gamma}(x_3)\cos \phi \\
\Delta_{a \gamma}(x_3)\sin \phi~  &  \Delta_{a \gamma}(x_3)\cos \phi~ & \Delta_{aa}(E)
\end{array}
\right)\, .
\end{eqnarray}

\begin{figure*}[t]
\centering
\subfigcapskip=0pt
\subfigbottomskip=0pt
\subfigure[{$m_a=10^{-10}\, \rm eV$, $g_{a\gamma}=10^{-9}\, \rm GeV^{-1}$.}]{\includegraphics[width=7.8cm]{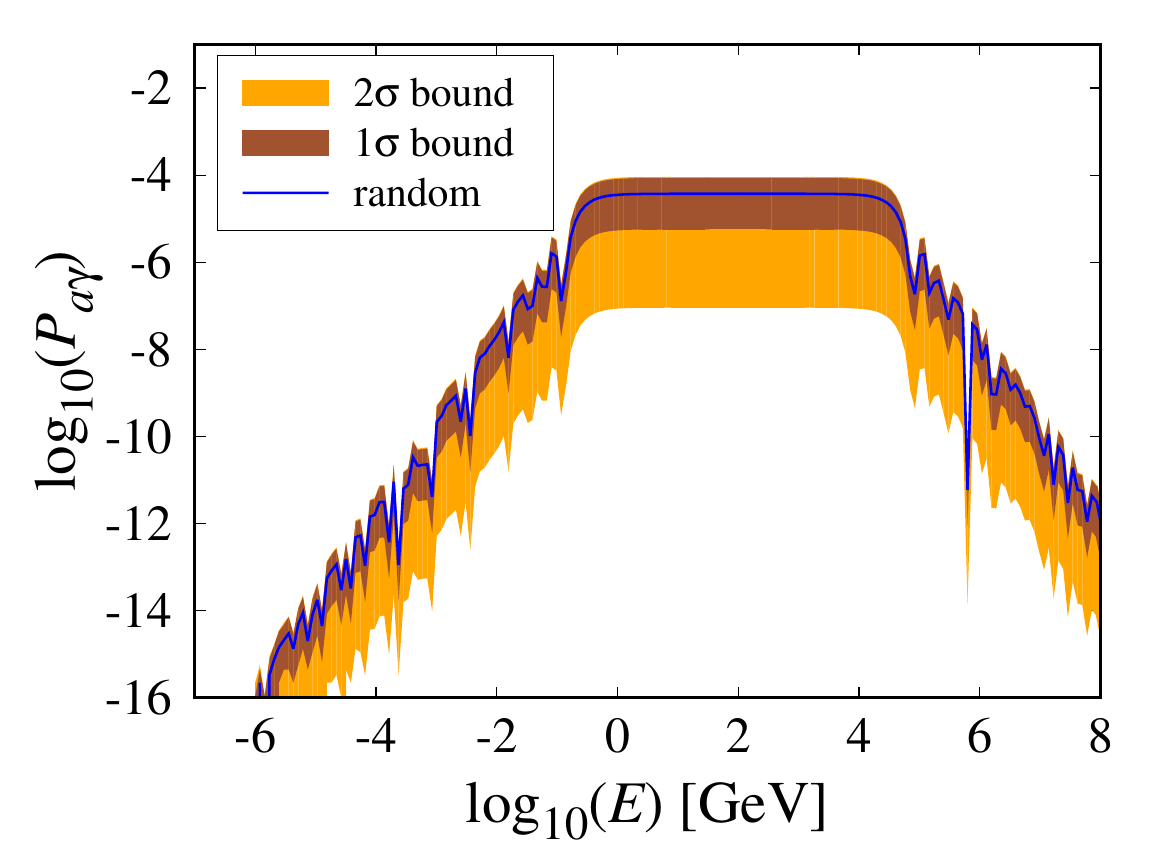}}\subfigure[{$m_a=10^{-8}\, \rm eV$, $g_{a\gamma}=10^{-9}\, \rm GeV^{-1}$.}]{\includegraphics[width=7.8cm]{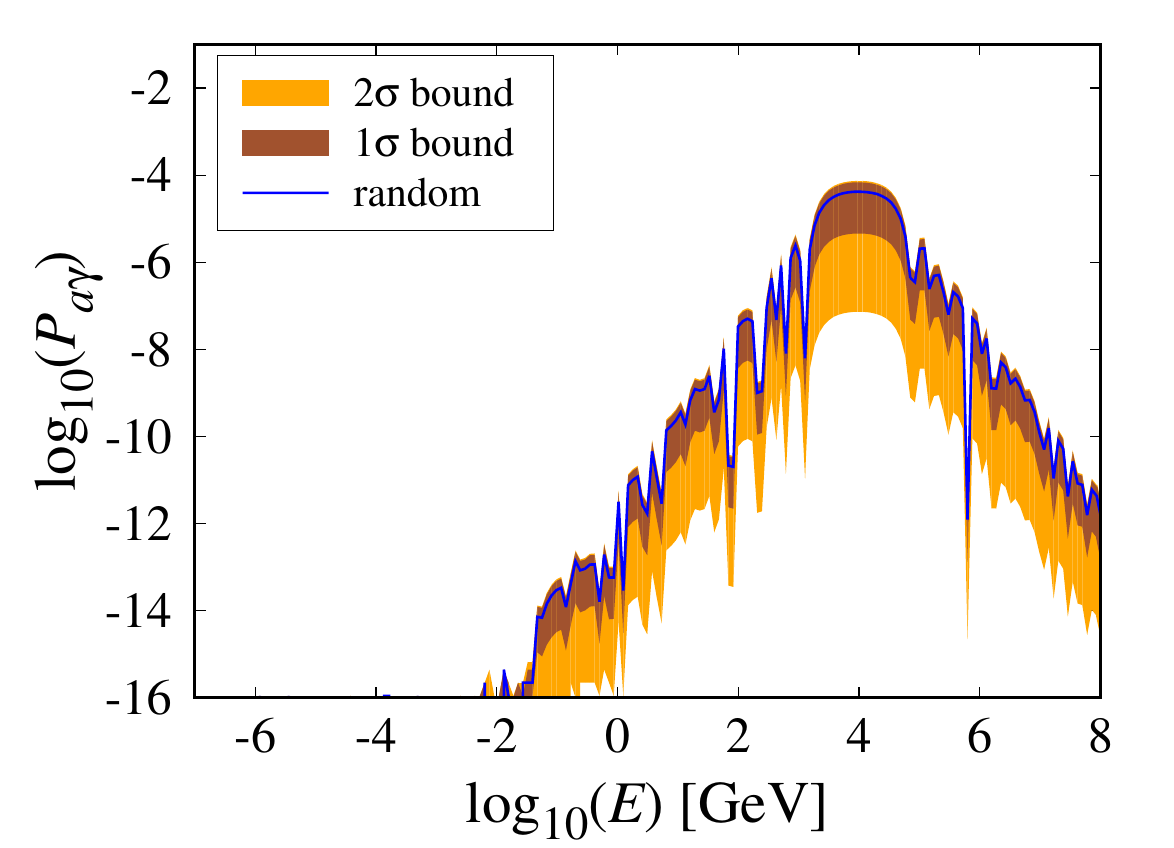}}
\subfigure[{$m_a=10^{-10}\, \rm eV$, $g_{a\gamma}=10^{-10}\, \rm GeV^{-1}$.}]{\includegraphics[width=7.8cm]{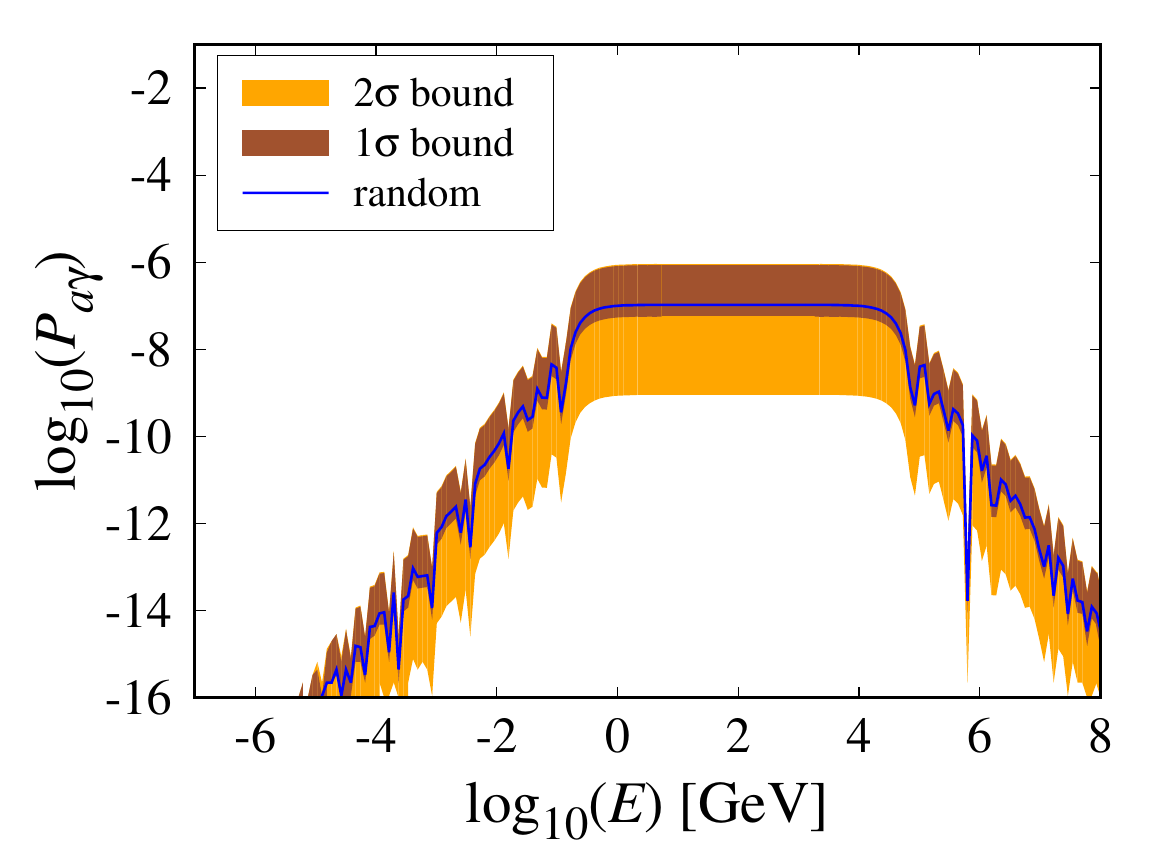}}\subfigure[{$m_a=10^{-8}\, \rm eV$, $g_{a\gamma}=10^{-10}\, \rm GeV^{-1}$.}]{\includegraphics[width=7.8cm]{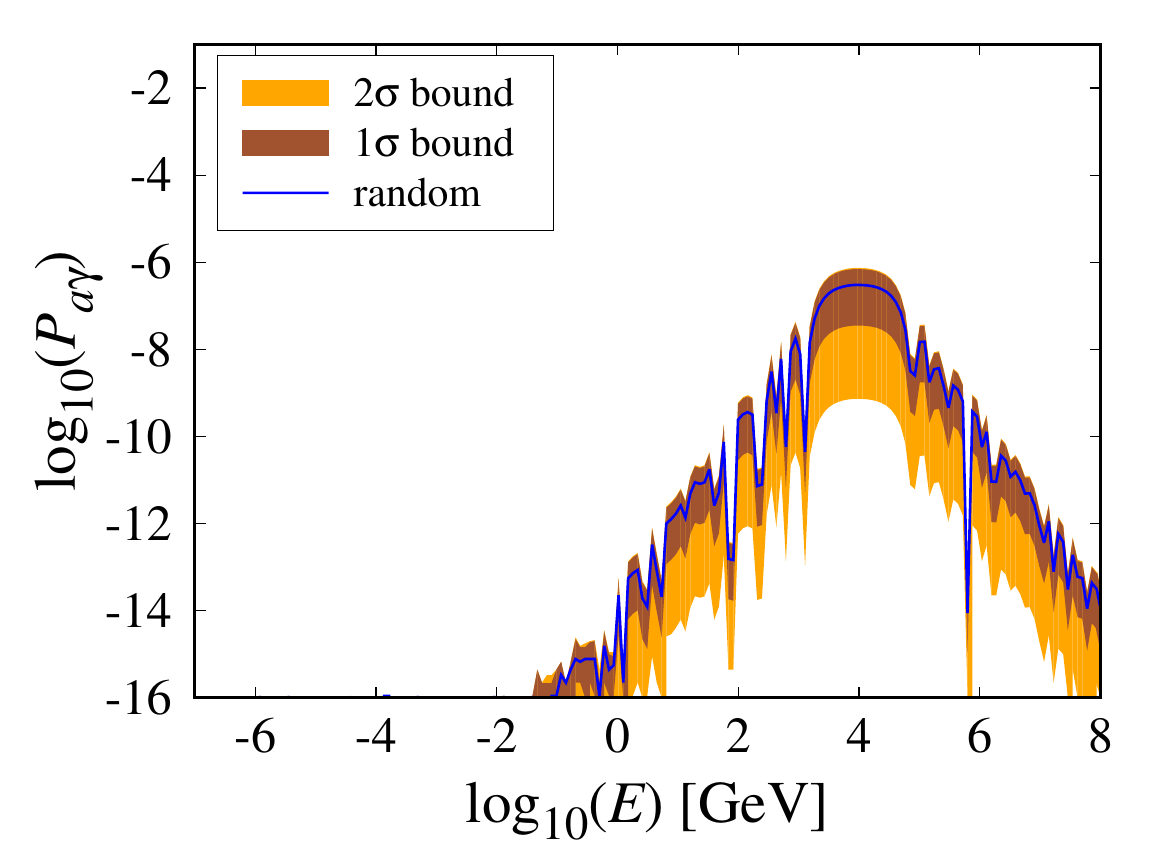}}
\caption{
ALP-photon oscillation probability in the stochastic magnetic field as a function of energy with the limit $B<8.9\, \rm pG$. Several typical ALP parameter sets [$m_a=10^{-10}\, \rm eV$, $g_{a\gamma}=10^{-9}\, \rm GeV^{-1}$] (a), [$m_a=10^{-8}\, \rm eV$, $g_{a\gamma}=10^{-9}\, \rm GeV^{-1}$] (b), [$m_a=10^{-10}\, \rm eV$, $g_{a\gamma}=10^{-10}\, \rm GeV^{-1}$] (c), and [$m_a=10^{-8}\, \rm eV$, $g_{a\gamma}=10^{-10}\, \rm GeV^{-1}$] (d) are selected for comparisons. The brown and orange bounds represent the $1\sigma$ and $2\sigma$ contours with 1000 random magnetic field configurations, respectively. The blue lines represent the oscillation probability with the random one configuration.
}
\label{fig_pa_ran}
\end{figure*}

Here we show the numerical results for the stochastic magnetic field scenario.
In this case, the magnetic field strength in every domain is same, while the direction is random from the domain $i$ to domain $i+1$, which corresponds to the PMF limit $B<8.9\, \rm pG$ with $n=5$.
In figure~\ref{fig_b_ran}, we plot the effective magnetic field strength distributions with this PMF limit.
The stochastic magnetic field is simulated with 100 domains for $1\, \rm Mpc$. 
The blue and red lines represent the effective magnetic field strength in each domain with the two stochastic magnetic field configurations. 
The dot-dashed black lines represent the magnetic field strength $|B_T| = 8.9\, \rm pG$.

In figure~\ref{fig_pa_ran}, we show the ALP-photon oscillation probability as a function of energy for the stochastic magnetic field scenario.
We show the several typical ALP parameter sets [$m_a=10^{-10}\, \rm eV$, $g_{a\gamma}=10^{-9}\, \rm GeV^{-1}$], [$m_a=10^{-8}\, \rm eV$, $g_{a\gamma}=10^{-9}\, \rm GeV^{-1}$], [$m_a=10^{-10}\, \rm eV$, $g_{a\gamma}=10^{-10}\, \rm GeV^{-1}$], and [$m_a=10^{-8}\, \rm eV$, $g_{a\gamma}=10^{-10}\, \rm GeV^{-1}$] for comparisons.
Here the magnetic field is simulated with 1000 random magnetic field configurations with the JS~2019 limit $B<8.9\, \rm pG$.
The brown and orange bounds represent the oscillation probability distributions with the $1\sigma$ and $2\sigma$ contours, respectively. 
The blue lines represent the oscillation probability with the random one magnetic field configuration.

As we can see, the ALP-photon oscillation probability for these typical ALP parameter sets show the different distributions.
For the same ALP mass $m_a$, the oscillation probability can be dramatically enhanced by the coupling $g_{a\gamma}$.
Compared with $g_{a\gamma}$, the mass $m_a$ can affect the oscillation probability in the low energy regions, while the probability would be stable for the high energies, which can also be seen from figure~\ref{fig_pa_contour_2}.
Note that the stochastic magnetic field is simulated with the PMF limit $B<8.9\, \rm pG$, the effect of magnetic field strength on ALP-photon oscillation probability can be found from the homogeneous magnetic field scenario as show in figures~\ref{fig_pa_contour_1} and \ref{fig_pa_contour_2}.

\subsection{Other effects on oscillations}

In order to obtain the current oscillation probability, the ALP-photon oscillation in the Universe evolution process may be affected by some other effects, such as the redshift and the photon absorption at the recombination \cite{Evoli:2016zhj, Schiavone:2021imu}.
Since we are more concerned about the phenomena of ALP-photon oscillations in the PMF, we just present these effects in this subsection and do not intend to show them in our numerical results. 
  
Considering the expansion of the Universe, we have the notations $B_T \to B_{T}\left(1+z\right)^2$, $E\to E\left(1+z\right)$, $l' \to l'\left(1+z\right)^{-1}$, $n_e\to n_{e}\left(1+z\right)^3$, $\Delta_{\rm pl}\to \Delta_{\rm pl}\left(1+z\right)^2$, $\Delta_{a\gamma}\to\Delta_{a\gamma}\left(1+z\right)^2$, $\Delta_{aa}\to\Delta_{aa}\left(1+z\right)^{-1}$, $\Delta_{\rm QED}\to\Delta_{\rm QED}\left(1+z\right)^5$, and $\Delta_{\rm CMB}\to\Delta_{\rm CMB}\left(1+z\right)^5$, where $z$ is the redshift, and $l'$ is the length of the magnetic field domain.

Additionally, the photon absorption by hydrogen and helium atoms formed at the recombination should also be considered.
We can rewrite the mixing matrix $\mathcal{M}(E,x_3)$ in eq.~(\ref{Mex3_2}) as
\begin{eqnarray}
\mathcal{M}(E,x_3)=
\left(
\begin{array}{ccc}
 \Delta_{11}(E,x_3)-i\frac{\Gamma(E)}{2}~ &  0&0 \\
 0& \Delta_{22}(E,x_3)-i\frac{\Gamma(E)}{2}~ & \Delta_{a \gamma}(x_3) \\
 0&\Delta_{a\gamma}(x_3)  & \Delta_{aa}(E)
\end{array}
\right)\, .
\end{eqnarray}
The term $\Gamma(E)$ represents the photon absorption rate, $\Gamma(E)=\sigma_{\rm H}(E)n_{\rm H}+\sigma_{\rm He}(E)n_{\rm He}+\sigma_{\rm KN}(E)n_{e}\, ,$ with the photoelectric cross-sections of Hydrogen $\sigma_{\rm H}$ and Helium $\sigma_{\rm He}$, and the Klein-Nishina cross-section for the Compton effect on electron $\sigma_{\rm KN}$. 
The number densities of $\rm H$, $\rm H_e$, and $e$ are $n_{\rm H}$, $n_{\rm He}$, and $n_{e}$, respectively. 
  
Considering the above effects on ALP-photon oscillations, one could derive the present-day oscillation probability $P'_{a\gamma}$ and the present-day photon flux 
\begin{eqnarray}
\frac{{\rm d}n'_\gamma(E,t_0)}{{\rm d}E}=P'_{a\gamma}\frac{{\rm d}n'_{\rm ALP}(E,t_0)}{{\rm d}E}\, .
\end{eqnarray}
In this case, the ALP-photon oscillations induced by PBHs may have the effects on some further phenomena for the photons in the different energy regions, such as the CMB, the CXB \cite{Schiavone:2021imu}, and the EGB. 
  
\section{Conclusions}
\label{sec_conclusions}

In this paper, we have presented the stochastic ALP-photon oscillations induced by PBHs in the PMF with the latest magnetic field limits.
First we introduce the PBH formation and evaporation, including the Schwarzschild black holes and Kerr black holes.
We investigate the ALPs emitted by ultra-light PBHs with the mass range $10 \, {\rm g} \lesssim M_{\rm PBH} \lesssim 10^9 \, \rm g$, in which PBHs would have completely evaporated before the start of BBN and can therefore not be directly constrained.
We show the instantaneous spectra of the spin-0 particle emitted by Schwarzschild and Kerr PBHs with the public code {\tt BlackHawk}.
The minimal scenario that ALPs could interact only with photons is supposed.
Then we study the stochastic oscillations between ALPs and photons in the cosmic magnetic field.

We discuss the ALP-photon oscillations in the PMF in detail, focusing on the distributions of oscillation probability with the homogeneous and stochastic magnetic field scenarios.
For the PMF, we consider the stochastic background field model as the completely non-homogeneous component of the cosmic plasma.
The stringent PMF limits are taken from JS~2019, which are shown as the scale-invariant ($n=0$, $B<47\, \rm pG$) and a violet Batchelor spectrum ($n=5$, $B<8.9\, \rm pG$) PMFs, correspond to the magnetic field strength of the homogeneous and stochastic magnetic field scenarios, respectively.
Above limits are derived from the CMB radiation based on the non-observation of the magnetic field induced clumping of baryonic density fluctuations during recombination epoch. 
For the homogeneous magnetic field, we show the ALP-photon oscillation probability distributions in the ALP parameter ($m_a$, $g_{a\gamma}$) space for several typical values of energy with the magnetic field strength $B_T=47\,\rm pG$ (JS~2019) and $\sim5\,\rm nG$ (Planck~2015).
The oscillation probability distribution changes significantly with the energy and magnetic field strength, and can be dramatically suppressed in the energy above $\sim10^4\, \rm GeV$. 
For the stochastic magnetic field, the magnetic field strength in every domain is supposed to be same, while the direction is random from the domain $i$ to $i+1$.
We show the effective magnetic field strength distributions with the JS~2019 PMF limit $B<8.9\, \rm pG$, which is simulated with 100 domains for $1\, \rm Mpc$.
We also show the numerical results of oscillation probability for the stochastic magnetic field, which is simulated with 1000 random magnetic field configurations.
The oscillation probability can be dramatically enhanced by the coupling $g_{a\gamma}$.
On the other hand, the oscillation probability can be affected by the ALP mass $m_a$ in the low energy regions and becomes stable for the high energies.
Finally, we briefly introduce the effects of redshift and photon absorption at the recombination on ALP-photon oscillations in the Universe evolution process.

In addition, the stochastic ALP-photon oscillations induced by PBHs may have the effects on some further phenomena, such as the CMB, the CXB, and the EGB, which may be detected as the signals of ALP-photon oscillations in the next generation ALP/photon measurements, $\rm e.g.$, ADMX \cite{ADMX:2018gho, ADMX:2019uok}, IAXO \cite{Armengaud:2014gea, IAXO:2019mpb}, ALPS-II \cite{Bahre:2013ywa, Spector:2016vwo}, LiteBIRD \cite{Matsumura:2013aja, Hazumi:2019lys}, CMB-S4 \cite{CMB-S4:2016ple, Abazajian:2019eic}, THESEUS \cite{THESEUS:2017qvx, THESEUS:2017wvz}, eXTP \cite{eXTP:2016rzs, eXTP:2018anb}, LHAASO \cite{Cao:2010zz, LHAASO:2019qtb}, and CTA \cite{CTAConsortium:2010umy, CTAConsortium:2013ofs}.
The stochastic ALP-photon oscillations in the PMF discussed in this paper may be of great importance to these topics.
     
\section*{Acknowledgments}
  
The author would like to thank the anonymous referee for valuable comments and suggestions improving the manuscript.
We also thank J\'er\'emy Auffinger, Junsong Cang, and Wei Chao for helpful discussions.
This work is supported by the National Natural Science Foundation (NNSF) of China (Grants No.~11775025 and No.~12175027).
    
   
\bibliographystyle{JHEP}
\bibliography{references}

\providecommand{\href}[2]{#2}\begingroup\raggedright\begin{thebibliography}{100}

\bibitem{Bird:2016dcv}
S.~Bird, I.~Cholis, J.~B. Mu\~noz, Y.~Ali-Ha\"\i{}moud, M.~Kamionkowski, E.~D.
  Kovetz et~al., \emph{{Did LIGO detect dark matter?}},
  \href{http://dx.doi.org/10.1103/PhysRevLett.116.201301}{\emph{Phys. Rev.
  Lett.} {\bf 116} (2016) 201301}, [\href{http://arxiv.org/abs/1603.00464}{{\tt
  1603.00464}}].

\bibitem{Hawking:1971ei}
S.~Hawking, \emph{{Gravitationally collapsed objects of very low mass}},
  {\emph{Mon. Not. Roy. Astron. Soc.} {\bf 152} (1971) 75}.

\bibitem{Carr:1974nx}
B.~J. Carr and S.~W. Hawking, \emph{{Black holes in the early Universe}},
  {\emph{Mon. Not. Roy. Astron. Soc.} {\bf 168} (1974) 399--415}.

\bibitem{Carr:1975qj}
B.~J. Carr, \emph{{The Primordial black hole mass spectrum}},
  \href{http://dx.doi.org/10.1086/153853}{\emph{Astrophys. J.} {\bf 201} (1975)
  1--19}.

\bibitem{Carr:2020gox}
B.~Carr, K.~Kohri, Y.~Sendouda and J.~Yokoyama, \emph{{Constraints on
  primordial black holes}},
  \href{http://dx.doi.org/10.1088/1361-6633/ac1e31}{\emph{Rept. Prog. Phys.}
  {\bf 84} (2021) 116902}, [\href{http://arxiv.org/abs/2002.12778}{{\tt
  2002.12778}}].

\bibitem{Carr:2020xqk}
B.~Carr and F.~Kuhnel, \emph{{Primordial Black Holes as Dark Matter: Recent
  Developments}},
  \href{http://dx.doi.org/10.1146/annurev-nucl-050520-125911}{\emph{Ann. Rev.
  Nucl. Part. Sci.} {\bf 70} (2020) 355--394},
  [\href{http://arxiv.org/abs/2006.02838}{{\tt 2006.02838}}].

\bibitem{Carr:2021bzv}
B.~Carr and F.~Kuhnel, \emph{{Primordial black holes as dark matter
  candidates}},
  \href{http://dx.doi.org/10.21468/SciPostPhysLectNotes.48}{\emph{SciPost Phys.
  Lect. Notes} {\bf 48} (2022) 1}, [\href{http://arxiv.org/abs/2110.02821}{{\tt
  2110.02821}}].

\bibitem{Carr:2016drx}
B.~Carr, F.~Kuhnel and M.~Sandstad, \emph{{Primordial Black Holes as Dark
  Matter}}, \href{http://dx.doi.org/10.1103/PhysRevD.94.083504}{\emph{Phys.
  Rev. D} {\bf 94} (2016) 083504}, [\href{http://arxiv.org/abs/1607.06077}{{\tt
  1607.06077}}].

\bibitem{Chapline:1975ojl}
G.~F. Chapline, \emph{{Cosmological effects of primordial black holes}},
  \href{http://dx.doi.org/10.1038/253251a0}{\emph{Nature} {\bf 253} (1975)
  251--252}.

\bibitem{Carr:2009jm}
B.~J. Carr, K.~Kohri, Y.~Sendouda and J.~Yokoyama, \emph{{New cosmological
  constraints on primordial black holes}},
  \href{http://dx.doi.org/10.1103/PhysRevD.81.104019}{\emph{Phys. Rev. D} {\bf
  81} (2010) 104019}, [\href{http://arxiv.org/abs/0912.5297}{{\tt 0912.5297}}].

\bibitem{Strong:2004ry}
A.~W. Strong, I.~V. Moskalenko and O.~Reimer, \emph{{A new determination of the
  extragalactic diffuse gamma-ray background from egret data}},
  \href{http://dx.doi.org/10.1086/423196}{\emph{Astrophys. J.} {\bf 613} (2004)
  956--961}, [\href{http://arxiv.org/abs/astro-ph/0405441}{{\tt
  astro-ph/0405441}}].

\bibitem{Barnacka:2012bm}
A.~Barnacka, J.~F. Glicenstein and R.~Moderski, \emph{{New constraints on
  primordial black holes abundance from femtolensing of gamma-ray bursts}},
  \href{http://dx.doi.org/10.1103/PhysRevD.86.043001}{\emph{Phys. Rev. D} {\bf
  86} (2012) 043001}, [\href{http://arxiv.org/abs/1204.2056}{{\tt 1204.2056}}].

\bibitem{Niikura:2019kqi}
H.~Niikura, M.~Takada, S.~Yokoyama, T.~Sumi and S.~Masaki, \emph{{Constraints
  on Earth-mass primordial black holes from OGLE 5-year microlensing events}},
  \href{http://dx.doi.org/10.1103/PhysRevD.99.083503}{\emph{Phys. Rev. D} {\bf
  99} (2019) 083503}, [\href{http://arxiv.org/abs/1901.07120}{{\tt
  1901.07120}}].

\bibitem{Carr:1997cn}
B.~J. Carr and M.~Sakellariadou, \emph{{Dynamical constraints on dark compact
  objects}}, \href{http://dx.doi.org/10.1086/307071}{\emph{Astrophys. J.} {\bf
  516} (1999) 195--220}.

\bibitem{Koushiappas:2017chw}
S.~M. Koushiappas and A.~Loeb, \emph{{Dynamics of Dwarf Galaxies Disfavor
  Stellar-Mass Black Holes as Dark Matter}},
  \href{http://dx.doi.org/10.1103/PhysRevLett.119.041102}{\emph{Phys. Rev.
  Lett.} {\bf 119} (2017) 041102}, [\href{http://arxiv.org/abs/1704.01668}{{\tt
  1704.01668}}].

\bibitem{Carr:2018rid}
B.~Carr and J.~Silk, \emph{{Primordial Black Holes as Generators of Cosmic
  Structures}}, \href{http://dx.doi.org/10.1093/mnras/sty1204}{\emph{Mon. Not.
  Roy. Astron. Soc.} {\bf 478} (2018) 3756--3775},
  [\href{http://arxiv.org/abs/1801.00672}{{\tt 1801.00672}}].

\bibitem{Afshordi:2003zb}
N.~Afshordi, P.~McDonald and D.~N. Spergel, \emph{{Primordial black holes as
  dark matter: The Power spectrum and evaporation of early structures}},
  \href{http://dx.doi.org/10.1086/378763}{\emph{Astrophys. J. Lett.} {\bf 594}
  (2003) L71--L74}, [\href{http://arxiv.org/abs/astro-ph/0302035}{{\tt
  astro-ph/0302035}}].

\bibitem{Raidal:2017mfl}
M.~Raidal, V.~Vaskonen and H.~Veerm\"ae, \emph{{Gravitational Waves from
  Primordial Black Hole Mergers}},
  \href{http://dx.doi.org/10.1088/1475-7516/2017/09/037}{\emph{JCAP} {\bf 09}
  (2017) 037}, [\href{http://arxiv.org/abs/1707.01480}{{\tt 1707.01480}}].

\bibitem{Hooper:2020evu}
D.~Hooper, G.~Krnjaic, J.~March-Russell, S.~D. McDermott and
  R.~Petrossian-Byrne, \emph{{Hot Gravitons and Gravitational Waves From Kerr
  Black Holes in the Early Universe}},
  \href{http://arxiv.org/abs/2004.00618}{{\tt 2004.00618}}.

\bibitem{Kohri:1999ex}
K.~Kohri and J.~Yokoyama, \emph{{Primordial black holes and primordial
  nucleosynthesis. 1. Effects of hadron injection from low mass holes}},
  \href{http://dx.doi.org/10.1103/PhysRevD.61.023501}{\emph{Phys. Rev. D} {\bf
  61} (2000) 023501}, [\href{http://arxiv.org/abs/astro-ph/9908160}{{\tt
  astro-ph/9908160}}].

\bibitem{Ricotti:2007au}
M.~Ricotti, J.~P. Ostriker and K.~J. Mack, \emph{{Effect of Primordial Black
  Holes on the Cosmic Microwave Background and Cosmological Parameter
  Estimates}}, \href{http://dx.doi.org/10.1086/587831}{\emph{Astrophys. J.}
  {\bf 680} (2008) 829}, [\href{http://arxiv.org/abs/0709.0524}{{\tt
  0709.0524}}].

\bibitem{Ali-Haimoud:2016mbv}
Y.~Ali-Ha\"\i{}moud and M.~Kamionkowski, \emph{{Cosmic microwave background
  limits on accreting primordial black holes}},
  \href{http://dx.doi.org/10.1103/PhysRevD.95.043534}{\emph{Phys. Rev. D} {\bf
  95} (2017) 043534}, [\href{http://arxiv.org/abs/1612.05644}{{\tt
  1612.05644}}].

\bibitem{Cang:2020aoo}
J.~Cang, Y.~Gao and Y.~Ma, \emph{{Prospects of Future CMB Anisotropy Probes for
  Primordial Black Holes}},
  \href{http://dx.doi.org/10.1088/1475-7516/2021/05/051}{\emph{JCAP} {\bf 05}
  (2021) 051}, [\href{http://arxiv.org/abs/2011.12244}{{\tt 2011.12244}}].

\bibitem{Maki:1995pa}
K.~Maki, T.~Mitsui and S.~Orito, \emph{{Local flux of low-energy anti-protons
  from evaporating primordial black holes}},
  \href{http://dx.doi.org/10.1103/PhysRevLett.76.3474}{\emph{Phys. Rev. Lett.}
  {\bf 76} (1996) 3474--3477},
  [\href{http://arxiv.org/abs/astro-ph/9601025}{{\tt astro-ph/9601025}}].

\bibitem{Linton:2006yu}
E.~T. Linton et~al., \emph{{A new search for primordial black hole evaporations
  using the Whipple gamma-ray telescope}},
  \href{http://dx.doi.org/10.1088/1475-7516/2006/01/013}{\emph{JCAP} {\bf 01}
  (2006) 013}.

\bibitem{Page:1976wx}
D.~N. Page and S.~W. Hawking, \emph{{Gamma rays from primordial black holes}},
  \href{http://dx.doi.org/10.1086/154350}{\emph{Astrophys. J.} {\bf 206} (1976)
  1--7}.

\bibitem{Carr:2016hva}
B.~J. Carr, K.~Kohri, Y.~Sendouda and J.~Yokoyama, \emph{{Constraints on
  primordial black holes from the Galactic gamma-ray background}},
  \href{http://dx.doi.org/10.1103/PhysRevD.94.044029}{\emph{Phys. Rev. D} {\bf
  94} (2016) 044029}, [\href{http://arxiv.org/abs/1604.05349}{{\tt
  1604.05349}}].

\bibitem{MacGibbon:1991vc}
J.~H. MacGibbon and B.~J. Carr, \emph{{Cosmic rays from primordial black
  holes}}, \href{http://dx.doi.org/10.1086/169909}{\emph{Astrophys. J.} {\bf
  371} (1991) 447--469}.

\bibitem{Bambi:2008kx}
C.~Bambi, A.~D. Dolgov and A.~A. Petrov, \emph{{Primordial black holes and the
  observed Galactic 511-keV line}},
  \href{http://dx.doi.org/10.1016/j.physletb.2009.10.053}{\emph{Phys. Lett. B}
  {\bf 670} (2008) 174--178}, [\href{http://arxiv.org/abs/0801.2786}{{\tt
  0801.2786}}]. [Erratum: Phys.Lett.B 681, 504--504 (2009)].

\bibitem{Lunardini:2019zob}
C.~Lunardini and Y.~F. Perez-Gonzalez, \emph{{Dirac and Majorana neutrino
  signatures of primordial black holes}},
  \href{http://dx.doi.org/10.1088/1475-7516/2020/08/014}{\emph{JCAP} {\bf 08}
  (2020) 014}, [\href{http://arxiv.org/abs/1910.07864}{{\tt 1910.07864}}].

\bibitem{Cang:2021owu}
J.~Cang, Y.~Gao and Y.-Z. Ma, \emph{{21-cm constraints on spinning primordial
  black holes}},
  \href{http://dx.doi.org/10.1088/1475-7516/2022/03/012}{\emph{JCAP} {\bf 03}
  (2022) 012}, [\href{http://arxiv.org/abs/2108.13256}{{\tt 2108.13256}}].

\bibitem{Dror:2021nyr}
J.~A. Dror, H.~Murayama and N.~L. Rodd, \emph{{Cosmic axion background}},
  \href{http://dx.doi.org/10.1103/PhysRevD.103.115004}{\emph{Phys. Rev. D} {\bf
  103} (2021) 115004}, [\href{http://arxiv.org/abs/2101.09287}{{\tt
  2101.09287}}].

\bibitem{Preskill:1982cy}
J.~Preskill, M.~B. Wise and F.~Wilczek, \emph{{Cosmology of the Invisible
  Axion}}, \href{http://dx.doi.org/10.1016/0370-2693(83)90637-8}{\emph{Phys.
  Lett. B} {\bf 120} (1983) 127--132}.

\bibitem{Sikivie:2009fv}
P.~Sikivie, \emph{{Dark matter axions}},
  \href{http://dx.doi.org/10.1142/S0217751X10048846}{\emph{Int. J. Mod. Phys.
  A} {\bf 25} (2010) 554--563}, [\href{http://arxiv.org/abs/0909.0949}{{\tt
  0909.0949}}].

\bibitem{Marsh:2015xka}
D.~J.~E. Marsh, \emph{{Axion Cosmology}},
  \href{http://dx.doi.org/10.1016/j.physrep.2016.06.005}{\emph{Phys. Rept.}
  {\bf 643} (2016) 1--79}, [\href{http://arxiv.org/abs/1510.07633}{{\tt
  1510.07633}}].

\bibitem{Bernal:2021yyb}
N.~Bernal, F.~Hajkarim and Y.~Xu, \emph{{Axion Dark Matter in the Time of
  Primordial Black Holes}},
  \href{http://dx.doi.org/10.1103/PhysRevD.104.075007}{\emph{Phys. Rev. D} {\bf
  104} (2021) 075007}, [\href{http://arxiv.org/abs/2107.13575}{{\tt
  2107.13575}}].

\bibitem{Bernal:2021bbv}
N.~Bernal, Y.~F. Perez-Gonzalez, Y.~Xu and O.~Zapata, \emph{{ALP dark matter in
  a primordial black hole dominated universe}},
  \href{http://dx.doi.org/10.1103/PhysRevD.104.123536}{\emph{Phys. Rev. D} {\bf
  104} (2021) 123536}, [\href{http://arxiv.org/abs/2110.04312}{{\tt
  2110.04312}}].

\bibitem{Choi:2022btl}
G.~Choi and E.~D. Schiappacasse, \emph{{PBH assisted search for QCD axion dark
  matter}}, \href{http://dx.doi.org/10.1088/1475-7516/2022/09/072}{\emph{JCAP}
  {\bf 09} (2022) 072}, [\href{http://arxiv.org/abs/2205.02255}{{\tt
  2205.02255}}].

\bibitem{Peccei:1977ur}
R.~Peccei and H.~R. Quinn, \emph{{Constraints Imposed by CP Conservation in the
  Presence of Instantons}},
  \href{http://dx.doi.org/10.1103/PhysRevD.16.1791}{\emph{Phys. Rev. D} {\bf
  16} (1977) 1791--1797}.

\bibitem{Peccei:1977hh}
R.~Peccei and H.~R. Quinn, \emph{{CP Conservation in the Presence of
  Instantons}},
  \href{http://dx.doi.org/10.1103/PhysRevLett.38.1440}{\emph{Phys. Rev. Lett.}
  {\bf 38} (1977) 1440--1443}.

\bibitem{Weinberg:1977ma}
S.~Weinberg, \emph{{A New Light Boson?}},
  \href{http://dx.doi.org/10.1103/PhysRevLett.40.223}{\emph{Phys. Rev. Lett.}
  {\bf 40} (1978) 223--226}.

\bibitem{Wilczek:1977pj}
F.~Wilczek, \emph{{Problem of Strong $P$ and $T$ Invariance in the Presence of
  Instantons}}, \href{http://dx.doi.org/10.1103/PhysRevLett.40.279}{\emph{Phys.
  Rev. Lett.} {\bf 40} (1978) 279--282}.

\bibitem{Svrcek:2006yi}
P.~Svrcek and E.~Witten, \emph{{Axions In String Theory}},
  \href{http://dx.doi.org/10.1088/1126-6708/2006/06/051}{\emph{JHEP} {\bf 06}
  (2006) 051}, [\href{http://arxiv.org/abs/hep-th/0605206}{{\tt
  hep-th/0605206}}].

\bibitem{Arvanitaki:2009fg}
A.~Arvanitaki, S.~Dimopoulos, S.~Dubovsky, N.~Kaloper and J.~March-Russell,
  \emph{{String Axiverse}},
  \href{http://dx.doi.org/10.1103/PhysRevD.81.123530}{\emph{Phys. Rev. D} {\bf
  81} (2010) 123530}, [\href{http://arxiv.org/abs/0905.4720}{{\tt 0905.4720}}].

\bibitem{DiLuzio:2020wdo}
L.~Di~Luzio, M.~Giannotti, E.~Nardi and L.~Visinelli, \emph{{The landscape of
  QCD axion models}},
  \href{http://dx.doi.org/10.1016/j.physrep.2020.06.002}{\emph{Phys. Rept.}
  {\bf 870} (2020) 1--117}, [\href{http://arxiv.org/abs/2003.01100}{{\tt
  2003.01100}}].

\bibitem{Choi:2020rgn}
K.~Choi, S.~H. Im and C.~Sub~Shin, \emph{{Recent Progress in the Physics of
  Axions and Axion-Like Particles}},
  \href{http://dx.doi.org/10.1146/annurev-nucl-120720-031147}{\emph{Ann. Rev.
  Nucl. Part. Sci.} {\bf 71} (2021) 225--252},
  [\href{http://arxiv.org/abs/2012.05029}{{\tt 2012.05029}}].

\bibitem{Galanti:2022ijh}
G.~Galanti and M.~Roncadelli, \emph{{Axion-like Particles Implications for
  High-Energy Astrophysics}},
  \href{http://dx.doi.org/10.3390/universe8050253}{\emph{Universe} {\bf 8}
  (2022) 253}, [\href{http://arxiv.org/abs/2205.00940}{{\tt 2205.00940}}].

\bibitem{Hooper:2007bq}
D.~Hooper and P.~D. Serpico, \emph{{Detecting Axion-Like Particles With Gamma
  Ray Telescopes}},
  \href{http://dx.doi.org/10.1103/PhysRevLett.99.231102}{\emph{Phys. Rev.
  Lett.} {\bf 99} (2007) 231102}, [\href{http://arxiv.org/abs/0706.3203}{{\tt
  0706.3203}}].

\bibitem{Li:2020pcn}
H.-J. Li, J.-G. Guo, X.-J. Bi, S.-J. Lin and P.-F. Yin, \emph{{Limits on
  axion-like particles from Mrk 421 with 4.5-year period observations by
  ARGO-YBJ and Fermi-LAT}},
  \href{http://dx.doi.org/10.1103/PhysRevD.103.083003}{\emph{Phys. Rev. D} {\bf
  103} (2021) 083003}, [\href{http://arxiv.org/abs/2008.09464}{{\tt
  2008.09464}}].

\bibitem{Li:2021zms}
H.-J. Li, \emph{{Relevance of VHE blazar spectra models with axion-like
  particles}},
  \href{http://dx.doi.org/10.1088/1475-7516/2022/02/025}{\emph{JCAP} {\bf 02}
  (2022) 025}, [\href{http://arxiv.org/abs/2112.14145}{{\tt 2112.14145}}].

\bibitem{Li:2022jgi}
H.-J. Li, \emph{{Probing photon-ALP oscillations from the flat spectrum radio
  quasar 4C+21.35}},
  \href{http://dx.doi.org/10.1016/j.physletb.2022.137047}{\emph{Phys. Lett. B}
  {\bf 829} (2022) 137047}, [\href{http://arxiv.org/abs/2203.08573}{{\tt
  2203.08573}}].

\bibitem{Giannotti:2015kwo}
M.~Giannotti, I.~Irastorza, J.~Redondo and A.~Ringwald, \emph{{Cool WISPs for
  stellar cooling excesses}},
  \href{http://dx.doi.org/10.1088/1475-7516/2016/05/057}{\emph{JCAP} {\bf 05}
  (2016) 057}, [\href{http://arxiv.org/abs/1512.08108}{{\tt 1512.08108}}].

\bibitem{Giannotti:2017hny}
M.~Giannotti, I.~G. Irastorza, J.~Redondo, A.~Ringwald and K.~Saikawa,
  \emph{{Stellar Recipes for Axion Hunters}},
  \href{http://dx.doi.org/10.1088/1475-7516/2017/10/010}{\emph{JCAP} {\bf 10}
  (2017) 010}, [\href{http://arxiv.org/abs/1708.02111}{{\tt 1708.02111}}].

\bibitem{Conlon:2013txa}
J.~P. Conlon and M.~C.~D. Marsh, \emph{{Excess Astrophysical Photons from a
  0.1\textendash{}1 keV Cosmic Axion Background}},
  \href{http://dx.doi.org/10.1103/PhysRevLett.111.151301}{\emph{Phys. Rev.
  Lett.} {\bf 111} (2013) 151301}, [\href{http://arxiv.org/abs/1305.3603}{{\tt
  1305.3603}}].

\bibitem{Angus:2013sua}
S.~Angus, J.~P. Conlon, M.~C.~D. Marsh, A.~J. Powell and L.~T. Witkowski,
  \emph{{Soft X-ray Excess in the Coma Cluster from a Cosmic Axion
  Background}},
  \href{http://dx.doi.org/10.1088/1475-7516/2014/09/026}{\emph{JCAP} {\bf 09}
  (2014) 026}, [\href{http://arxiv.org/abs/1312.3947}{{\tt 1312.3947}}].

\bibitem{XENON:2020rca}
{\scshape XENON} collaboration, E.~Aprile et~al., \emph{{Excess electronic
  recoil events in XENON1T}},
  \href{http://dx.doi.org/10.1103/PhysRevD.102.072004}{\emph{Phys. Rev. D} {\bf
  102} (2020) 072004}, [\href{http://arxiv.org/abs/2006.09721}{{\tt
  2006.09721}}].

\bibitem{DiLuzio:2020jjp}
L.~Di~Luzio, M.~Fedele, M.~Giannotti, F.~Mescia and E.~Nardi, \emph{{Solar
  axions cannot explain the XENON1T excess}},
  \href{http://dx.doi.org/10.1103/PhysRevLett.125.131804}{\emph{Phys. Rev.
  Lett.} {\bf 125} (2020) 131804}, [\href{http://arxiv.org/abs/2006.12487}{{\tt
  2006.12487}}].

\bibitem{ciaran_o_hare_2020_3932430}
C.~O'HARE, \emph{cajohare/axionlimits: Axionlimits},  July, 2020.
\newblock 10.5281/zenodo.3932430.

\bibitem{Subramanian:2015lua}
K.~Subramanian, \emph{{The origin, evolution and signatures of primordial
  magnetic fields}},
  \href{http://dx.doi.org/10.1088/0034-4885/79/7/076901}{\emph{Rept. Prog.
  Phys.} {\bf 79} (2016) 076901}, [\href{http://arxiv.org/abs/1504.02311}{{\tt
  1504.02311}}].

\bibitem{Turner:1987bw}
M.~S. Turner and L.~M. Widrow, \emph{{Inflation Produced, Large Scale Magnetic
  Fields}}, \href{http://dx.doi.org/10.1103/PhysRevD.37.2743}{\emph{Phys. Rev.
  D} {\bf 37} (1988) 2743}.

\bibitem{Ratra:1991bn}
B.~Ratra, \emph{{Cosmological 'seed' magnetic field from inflation}},
  \href{http://dx.doi.org/10.1086/186384}{\emph{Astrophys. J. Lett.} {\bf 391}
  (1992) L1--L4}.

\bibitem{Vachaspati:1991nm}
T.~Vachaspati, \emph{{Magnetic fields from cosmological phase transitions}},
  \href{http://dx.doi.org/10.1016/0370-2693(91)90051-Q}{\emph{Phys. Lett. B}
  {\bf 265} (1991) 258--261}.

\bibitem{Grasso:1997nx}
D.~Grasso and A.~Riotto, \emph{{On the nature of the magnetic fields generated
  during the electroweak phase transition}},
  \href{http://dx.doi.org/10.1016/S0370-2693(97)01224-0}{\emph{Phys. Lett. B}
  {\bf 418} (1998) 258--265}, [\href{http://arxiv.org/abs/hep-ph/9707265}{{\tt
  hep-ph/9707265}}].

\bibitem{Brandenburg:2017neh}
A.~Brandenburg, T.~Kahniashvili, S.~Mandal, A.~Roper~Pol, A.~G. Tevzadze and
  T.~Vachaspati, \emph{{Evolution of hydromagnetic turbulence from the
  electroweak phase transition}},
  \href{http://dx.doi.org/10.1103/PhysRevD.96.123528}{\emph{Phys. Rev. D} {\bf
  96} (2017) 123528}, [\href{http://arxiv.org/abs/1711.03804}{{\tt
  1711.03804}}].

\bibitem{Grasso:1994ph}
D.~Grasso and H.~R. Rubinstein, \emph{{Limits on possible magnetic fields at
  nucleosynthesis time}},
  \href{http://dx.doi.org/10.1016/0927-6505(94)00030-7}{\emph{Astropart. Phys.}
  {\bf 3} (1995) 95--102}, [\href{http://arxiv.org/abs/astro-ph/9409010}{{\tt
  astro-ph/9409010}}].

\bibitem{Kahniashvili:2010wm}
T.~Kahniashvili, A.~G. Tevzadze, S.~K. Sethi, K.~Pandey and B.~Ratra,
  \emph{{Primordial magnetic field limits from cosmological data}},
  \href{http://dx.doi.org/10.1103/PhysRevD.82.083005}{\emph{Phys. Rev. D} {\bf
  82} (2010) 083005}, [\href{http://arxiv.org/abs/1009.2094}{{\tt 1009.2094}}].

\bibitem{Jedamzik:1999bm}
K.~Jedamzik, V.~Katalinic and A.~V. Olinto, \emph{{A Limit on primordial small
  scale magnetic fields from CMB distortions}},
  \href{http://dx.doi.org/10.1103/PhysRevLett.85.700}{\emph{Phys. Rev. Lett.}
  {\bf 85} (2000) 700--703}, [\href{http://arxiv.org/abs/astro-ph/9911100}{{\tt
  astro-ph/9911100}}].

\bibitem{Kunze:2013uja}
K.~E. Kunze and E.~Komatsu, \emph{{Constraining primordial magnetic fields with
  distortions of the black-body spectrum of the cosmic microwave background:
  pre- and post-decoupling contributions}},
  \href{http://dx.doi.org/10.1088/1475-7516/2014/01/009}{\emph{JCAP} {\bf 01}
  (2014) 009}, [\href{http://arxiv.org/abs/1309.7994}{{\tt 1309.7994}}].

\bibitem{Kim:1994zh}
E.-j. Kim, A.~Olinto and R.~Rosner, \emph{{Generation of density perturbations
  by primordial magnetic fields}},
  \href{http://dx.doi.org/10.1086/177667}{\emph{Astrophys. J.} {\bf 468} (1996)
  28}, [\href{http://arxiv.org/abs/astro-ph/9412070}{{\tt astro-ph/9412070}}].

\bibitem{Battaner:1996jk}
E.~Battaner, E.~Florido and J.~Jimenez-Vicente, \emph{{Magnetic fields and
  large scale structure in a radiation dominated universe}}, {\emph{Astron.
  Astrophys.} {\bf 326} (1997) 13--22},
  [\href{http://arxiv.org/abs/astro-ph/9602097}{{\tt astro-ph/9602097}}].

\bibitem{Caprini:2001nb}
C.~Caprini and R.~Durrer, \emph{{Gravitational wave production: A Strong
  constraint on primordial magnetic fields}},
  \href{http://dx.doi.org/10.1103/PhysRevD.65.023517}{\emph{Phys. Rev. D} {\bf
  65} (2001) 023517}, [\href{http://arxiv.org/abs/astro-ph/0106244}{{\tt
  astro-ph/0106244}}].

\bibitem{Caprini:2006jb}
C.~Caprini and R.~Durrer, \emph{{Gravitational waves from stochastic
  relativistic sources: Primordial turbulence and magnetic fields}},
  \href{http://dx.doi.org/10.1103/PhysRevD.74.063521}{\emph{Phys. Rev. D} {\bf
  74} (2006) 063521}, [\href{http://arxiv.org/abs/astro-ph/0603476}{{\tt
  astro-ph/0603476}}].

\bibitem{RoperPol:2019wvy}
A.~Roper~Pol, S.~Mandal, A.~Brandenburg, T.~Kahniashvili and A.~Kosowsky,
  \emph{{Numerical simulations of gravitational waves from early-universe
  turbulence}},
  \href{http://dx.doi.org/10.1103/PhysRevD.102.083512}{\emph{Phys. Rev. D} {\bf
  102} (2020) 083512}, [\href{http://arxiv.org/abs/1903.08585}{{\tt
  1903.08585}}].

\bibitem{Yamazaki:2012pg}
D.~G. Yamazaki, T.~Kajino, G.~J. Mathew and K.~Ichiki, \emph{{The Search for a
  Primordial Magnetic Field}},
  \href{http://dx.doi.org/10.1016/j.physrep.2012.02.005}{\emph{Phys. Rept.}
  {\bf 517} (2012) 141--167}, [\href{http://arxiv.org/abs/1204.3669}{{\tt
  1204.3669}}].

\bibitem{Durrer:2013pga}
R.~Durrer and A.~Neronov, \emph{{Cosmological Magnetic Fields: Their
  Generation, Evolution and Observation}},
  \href{http://dx.doi.org/10.1007/s00159-013-0062-7}{\emph{Astron. Astrophys.
  Rev.} {\bf 21} (2013) 62}, [\href{http://arxiv.org/abs/1303.7121}{{\tt
  1303.7121}}].

\bibitem{Planck:2015zrl}
{\scshape Planck} collaboration, P.~A.~R. Ade et~al., \emph{{Planck 2015
  results. XIX. Constraints on primordial magnetic fields}},
  \href{http://dx.doi.org/10.1051/0004-6361/201525821}{\emph{Astron.
  Astrophys.} {\bf 594} (2016) A19},
  [\href{http://arxiv.org/abs/1502.01594}{{\tt 1502.01594}}].

\bibitem{Jedamzik:2018itu}
K.~Jedamzik and A.~Saveliev, \emph{{Stringent Limit on Primordial Magnetic
  Fields from the Cosmic Microwave Background Radiation}},
  \href{http://dx.doi.org/10.1103/PhysRevLett.123.021301}{\emph{Phys. Rev.
  Lett.} {\bf 123} (2019) 021301}, [\href{http://arxiv.org/abs/1804.06115}{{\tt
  1804.06115}}].

\bibitem{Schiavone:2021imu}
F.~Schiavone, D.~Montanino, A.~Mirizzi and F.~Capozzi, \emph{{Axion-like
  particles from primordial black holes shining through the Universe}},
  \href{http://dx.doi.org/10.1088/1475-7516/2021/08/063}{\emph{JCAP} {\bf 08}
  (2021) 063}, [\href{http://arxiv.org/abs/2107.03420}{{\tt 2107.03420}}].

\bibitem{Evoli:2016zhj}
C.~Evoli, M.~Leo, A.~Mirizzi and D.~Montanino, \emph{{Reionization during the
  dark ages from a cosmic axion background}},
  \href{http://dx.doi.org/10.1088/1475-7516/2016/05/006}{\emph{JCAP} {\bf 05}
  (2016) 006}, [\href{http://arxiv.org/abs/1602.08433}{{\tt 1602.08433}}].

\bibitem{Papanikolaou:2020qtd}
T.~Papanikolaou, V.~Vennin and D.~Langlois, \emph{{Gravitational waves from a
  universe filled with primordial black holes}},
  \href{http://dx.doi.org/10.1088/1475-7516/2021/03/053}{\emph{JCAP} {\bf 03}
  (2021) 053}, [\href{http://arxiv.org/abs/2010.11573}{{\tt 2010.11573}}].

\bibitem{Hawking:1974rv}
S.~W. Hawking, \emph{{Black hole explosions}},
  \href{http://dx.doi.org/10.1038/248030a0}{\emph{Nature} {\bf 248} (1974)
  30--31}.

\bibitem{Planck:2018jri}
{\scshape Planck} collaboration, Y.~Akrami et~al., \emph{{Planck 2018 results.
  X. Constraints on inflation}},
  \href{http://dx.doi.org/10.1051/0004-6361/201833887}{\emph{Astron.
  Astrophys.} {\bf 641} (2020) A10},
  [\href{http://arxiv.org/abs/1807.06211}{{\tt 1807.06211}}].

\bibitem{Hawking:1975vcx}
S.~W. Hawking, \emph{{Particle Creation by Black Holes}},
  \href{http://dx.doi.org/10.1007/BF02345020}{\emph{Commun. Math. Phys.} {\bf
  43} (1975) 199--220}. [Erratum: Commun.Math.Phys. 46, 206 (1976)].

\bibitem{Page:1976ki}
D.~N. Page, \emph{{Particle Emission Rates from a Black Hole. 2. Massless
  Particles from a Rotating Hole}},
  \href{http://dx.doi.org/10.1103/PhysRevD.14.3260}{\emph{Phys. Rev. D} {\bf
  14} (1976) 3260--3273}.

\bibitem{MacGibbon:1990zk}
J.~H. MacGibbon and B.~R. Webber, \emph{{Quark and gluon jet emission from
  primordial black holes: The instantaneous spectra}},
  \href{http://dx.doi.org/10.1103/PhysRevD.41.3052}{\emph{Phys. Rev. D} {\bf
  41} (1990) 3052--3079}.

\bibitem{Arbey:2019mbc}
A.~Arbey and J.~Auffinger, \emph{{BlackHawk: A public code for calculating the
  Hawking evaporation spectra of any black hole distribution}},
  \href{http://dx.doi.org/10.1140/epjc/s10052-019-7161-1}{\emph{Eur. Phys. J.
  C} {\bf 79} (2019) 693}, [\href{http://arxiv.org/abs/1905.04268}{{\tt
  1905.04268}}].

\bibitem{Arbey:2021mbl}
A.~Arbey and J.~Auffinger, \emph{{Physics Beyond the Standard Model with
  BlackHawk v2.0}},
  \href{http://dx.doi.org/10.1140/epjc/s10052-021-09702-8}{\emph{Eur. Phys. J.
  C} {\bf 81} (2021) 910}, [\href{http://arxiv.org/abs/2108.02737}{{\tt
  2108.02737}}].

\bibitem{Page:1974he}
D.~N. Page and K.~S. Thorne, \emph{{Disk-Accretion onto a Black Hole.
  Time-Averaged Structure of Accretion Disk}},
  \href{http://dx.doi.org/10.1086/152990}{\emph{Astrophys. J.} {\bf 191} (1974)
  499--506}.

\bibitem{Buonanno:2007sv}
A.~Buonanno, L.~E. Kidder and L.~Lehner, \emph{{Estimating the final spin of a
  binary black hole coalescence}},
  \href{http://dx.doi.org/10.1103/PhysRevD.77.026004}{\emph{Phys. Rev. D} {\bf
  77} (2008) 026004}, [\href{http://arxiv.org/abs/0709.3839}{{\tt 0709.3839}}].

\bibitem{Page:1976df}
D.~N. Page, \emph{{Particle Emission Rates from a Black Hole: Massless
  Particles from an Uncharged, Nonrotating Hole}},
  \href{http://dx.doi.org/10.1103/PhysRevD.13.198}{\emph{Phys. Rev. D} {\bf 13}
  (1976) 198--206}.

\bibitem{Barrow:1997mj}
J.~D. Barrow, P.~G. Ferreira and J.~Silk, \emph{{Constraints on a primordial
  magnetic field}},
  \href{http://dx.doi.org/10.1103/PhysRevLett.78.3610}{\emph{Phys. Rev. Lett.}
  {\bf 78} (1997) 3610--3613},
  [\href{http://arxiv.org/abs/astro-ph/9701063}{{\tt astro-ph/9701063}}].

\bibitem{Kahniashvili:2008sh}
T.~Kahniashvili, G.~Lavrelashvili and B.~Ratra, \emph{{CMB Temperature
  Anisotropy from Broken Spatial Isotropy due to an Homogeneous Cosmological
  Magnetic Field}},
  \href{http://dx.doi.org/10.1103/PhysRevD.78.063012}{\emph{Phys. Rev. D} {\bf
  78} (2008) 063012}, [\href{http://arxiv.org/abs/0807.4239}{{\tt 0807.4239}}].

\bibitem{Lewis:2004ef}
A.~Lewis, \emph{{CMB anisotropies from primordial inhomogeneous magnetic
  fields}}, \href{http://dx.doi.org/10.1103/PhysRevD.70.043011}{\emph{Phys.
  Rev. D} {\bf 70} (2004) 043011},
  [\href{http://arxiv.org/abs/astro-ph/0406096}{{\tt astro-ph/0406096}}].

\bibitem{Paoletti:2008ck}
D.~Paoletti, F.~Finelli and F.~Paci, \emph{{The full contribution of a
  stochastic background of magnetic fields to CMB anisotropies}},
  \href{http://dx.doi.org/10.1111/j.1365-2966.2009.14727.x}{\emph{Mon. Not.
  Roy. Astron. Soc.} {\bf 396} (2009) 523--534},
  [\href{http://arxiv.org/abs/0811.0230}{{\tt 0811.0230}}].

\bibitem{Mack:2001gc}
A.~Mack, T.~Kahniashvili and A.~Kosowsky, \emph{{Microwave background
  signatures of a primordial stochastic magnetic field}},
  \href{http://dx.doi.org/10.1103/PhysRevD.65.123004}{\emph{Phys. Rev. D} {\bf
  65} (2002) 123004}, [\href{http://arxiv.org/abs/astro-ph/0105504}{{\tt
  astro-ph/0105504}}].

\bibitem{Raffelt:1987im}
G.~Raffelt and L.~Stodolsky, \emph{{Mixing of the Photon with Low Mass
  Particles}}, \href{http://dx.doi.org/10.1103/PhysRevD.37.1237}{\emph{Phys.
  Rev. D} {\bf 37} (1988) 1237}.

\bibitem{DeAngelis:2011id}
A.~De~Angelis, G.~Galanti and M.~Roncadelli, \emph{{Relevance of axion-like
  particles for very-high-energy astrophysics}},
  \href{http://dx.doi.org/10.1103/PhysRevD.84.105030}{\emph{Phys. Rev. D} {\bf
  84} (2011) 105030}, [\href{http://arxiv.org/abs/1106.1132}{{\tt 1106.1132}}].
  [Erratum: Phys.Rev.D 87, 109903 (2013)].

\bibitem{Schwinger:1951nm}
J.~S. Schwinger, \emph{{On gauge invariance and vacuum polarization}},
  \href{http://dx.doi.org/10.1103/PhysRev.82.664}{\emph{Phys. Rev.} {\bf 82}
  (1951) 664--679}.

\bibitem{Dobrynina:2014qba}
A.~Dobrynina, A.~Kartavtsev and G.~Raffelt, \emph{{Photon-photon dispersion of
  TeV gamma rays and its role for photon-ALP conversion}},
  \href{http://dx.doi.org/10.1103/PhysRevD.91.083003}{\emph{Phys. Rev. D} {\bf
  91} (2015) 083003}, [\href{http://arxiv.org/abs/1412.4777}{{\tt 1412.4777}}].
  [Erratum: Phys.Rev.D 95, 109905 (2017)].

\bibitem{ADMX:2018gho}
{\scshape ADMX} collaboration, N.~Du et~al., \emph{{A Search for Invisible
  Axion Dark Matter with the Axion Dark Matter Experiment}},
  \href{http://dx.doi.org/10.1103/PhysRevLett.120.151301}{\emph{Phys. Rev.
  Lett.} {\bf 120} (2018) 151301}, [\href{http://arxiv.org/abs/1804.05750}{{\tt
  1804.05750}}].

\bibitem{ADMX:2019uok}
{\scshape ADMX} collaboration, T.~Braine et~al., \emph{{Extended Search for the
  Invisible Axion with the Axion Dark Matter Experiment}},
  \href{http://dx.doi.org/10.1103/PhysRevLett.124.101303}{\emph{Phys. Rev.
  Lett.} {\bf 124} (2020) 101303}, [\href{http://arxiv.org/abs/1910.08638}{{\tt
  1910.08638}}].

\bibitem{Armengaud:2014gea}
E.~Armengaud et~al., \emph{{Conceptual Design of the International Axion
  Observatory (IAXO)}},
  \href{http://dx.doi.org/10.1088/1748-0221/9/05/T05002}{\emph{JINST} {\bf 9}
  (2014) T05002}, [\href{http://arxiv.org/abs/1401.3233}{{\tt 1401.3233}}].

\bibitem{IAXO:2019mpb}
{\scshape IAXO} collaboration, E.~Armengaud et~al., \emph{{Physics potential of
  the International Axion Observatory (IAXO)}},
  \href{http://dx.doi.org/10.1088/1475-7516/2019/06/047}{\emph{JCAP} {\bf 06}
  (2019) 047}, [\href{http://arxiv.org/abs/1904.09155}{{\tt 1904.09155}}].

\bibitem{Bahre:2013ywa}
R.~B\"ahre et~al., \emph{{Any light particle search II \textemdash{}Technical
  Design Report}},
  \href{http://dx.doi.org/10.1088/1748-0221/8/09/T09001}{\emph{JINST} {\bf 8}
  (2013) T09001}, [\href{http://arxiv.org/abs/1302.5647}{{\tt 1302.5647}}].

\bibitem{Spector:2016vwo}
{\scshape ALPS} collaboration, A.~Spector, \emph{{ALPS II technical overview
  and status report}},  in \emph{{12th Patras Workshop on Axions, WIMPs and
  WISPs}}, pp.~133--136, 2017.
\newblock \href{http://arxiv.org/abs/1611.05863}{{\tt 1611.05863}}.
\newblock
  \href{http://dx.doi.org/10.3204/DESY-PROC-2009-03/Spector_Aaron}{DOI}.

\bibitem{Matsumura:2013aja}
T.~Matsumura et~al., \emph{{Mission design of LiteBIRD}},
  \href{http://dx.doi.org/10.1007/s10909-013-0996-1}{\emph{J. Low Temp. Phys.}
  {\bf 176} (2014) 733}, [\href{http://arxiv.org/abs/1311.2847}{{\tt
  1311.2847}}].

\bibitem{Hazumi:2019lys}
M.~Hazumi et~al., \emph{{LiteBIRD: A Satellite for the Studies of B-Mode
  Polarization and Inflation from Cosmic Background Radiation Detection}},
  \href{http://dx.doi.org/10.1007/s10909-019-02150-5}{\emph{J. Low Temp. Phys.}
  {\bf 194} (2019) 443--452}.

\bibitem{CMB-S4:2016ple}
{\scshape CMB-S4} collaboration, K.~N. Abazajian et~al., \emph{{CMB-S4 Science
  Book, First Edition}},  \href{http://arxiv.org/abs/1610.02743}{{\tt
  1610.02743}}.

\bibitem{Abazajian:2019eic}
K.~Abazajian et~al., \emph{{CMB-S4 Science Case, Reference Design, and Project
  Plan}},  \href{http://arxiv.org/abs/1907.04473}{{\tt 1907.04473}}.

\bibitem{THESEUS:2017qvx}
{\scshape THESEUS} collaboration, L.~Amati et~al., \emph{{The THESEUS space
  mission concept: science case, design and expected performances}},
  \href{http://dx.doi.org/10.1016/j.asr.2018.03.010}{\emph{Adv. Space Res.}
  {\bf 62} (2018) 191--244}, [\href{http://arxiv.org/abs/1710.04638}{{\tt
  1710.04638}}].

\bibitem{THESEUS:2017wvz}
{\scshape THESEUS} collaboration, G.~Stratta et~al., \emph{{THESEUS: a key
  space mission concept for Multi-Messenger Astrophysics}},
  \href{http://dx.doi.org/10.1016/j.asr.2018.04.013}{\emph{Adv. Space Res.}
  {\bf 62} (2018) 662--682}, [\href{http://arxiv.org/abs/1712.08153}{{\tt
  1712.08153}}].

\bibitem{eXTP:2016rzs}
{\scshape eXTP} collaboration, S.~N. Zhang et~al., \emph{{eXTP -- enhanced
  X-ray Timing and Polarimetry Mission}},
  \href{http://dx.doi.org/10.1117/12.2232034}{\emph{Proc. SPIE Int. Soc. Opt.
  Eng.} {\bf 9905} (2016) 99051Q}, [\href{http://arxiv.org/abs/1607.08823}{{\tt
  1607.08823}}].

\bibitem{eXTP:2018anb}
{\scshape eXTP} collaboration, S.-N. Zhang et~al., \emph{{The enhanced X-ray
  Timing and Polarimetry mission\textemdash{}eXTP}},
  \href{http://dx.doi.org/10.1007/s11433-018-9309-2}{\emph{Sci. China Phys.
  Mech. Astron.} {\bf 62} (2019) 29502},
  [\href{http://arxiv.org/abs/1812.04020}{{\tt 1812.04020}}].

\bibitem{Cao:2010zz}
{\scshape LHAASO} collaboration, Z.~Cao, \emph{{A future project at Tibet: The
  large high altitude air shower observatory (LHAASO)}},
  \href{http://dx.doi.org/10.1088/1674-1137/34/2/018}{\emph{Chin. Phys. C} {\bf
  34} (2010) 249--252}.

\bibitem{LHAASO:2019qtb}
{\scshape LHAASO} collaboration, A.~Addazi et~al., \emph{{The Large High
  Altitude Air Shower Observatory (LHAASO) Science Book (2021 Edition)}},
  {\emph{Chin. Phys. C} {\bf 46} (2022) 035001--035007},
  [\href{http://arxiv.org/abs/1905.02773}{{\tt 1905.02773}}].

\bibitem{CTAConsortium:2010umy}
{\scshape CTA Consortium} collaboration, M.~Actis et~al., \emph{{Design
  concepts for the Cherenkov Telescope Array CTA: An advanced facility for
  ground-based high-energy gamma-ray astronomy}},
  \href{http://dx.doi.org/10.1007/s10686-011-9247-0}{\emph{Exper. Astron.} {\bf
  32} (2011) 193--316}, [\href{http://arxiv.org/abs/1008.3703}{{\tt
  1008.3703}}].

\bibitem{CTAConsortium:2013ofs}
{\scshape CTA Consortium} collaboration, B.~S. Acharya et~al.,
  \emph{{Introducing the CTA concept}},
  \href{http://dx.doi.org/10.1016/j.astropartphys.2013.01.007}{\emph{Astropart.
  Phys.} {\bf 43} (2013) 3--18}.

\end{thebibliography}\endgroup

\end{document}